%% file: AnonymousSubmission2027.tex
\documentclass[letterpaper]{article} 
\usepackage[preprint]{aaai2027}  
\usepackage[hyphens]{url}  
\usepackage{graphicx} 
\usepackage{natbib}  
\usepackage{caption} 
\usepackage{algorithm}
\usepackage{algorithmic}

\usepackage{newfloat}
\usepackage{listings}
\DeclareCaptionStyle{ruled}{labelfont=normalfont,labelsep=colon,strut=off} 
\floatstyle{ruled}
\newfloat{listing}{tb}{lst}{}
\floatname{listing}{Listing}

\usepackage{booktabs}
\usepackage{multirow}
\usepackage{kotex}
\usepackage{amsmath}
\usepackage{amssymb}

\usepackage{array}
\title{LAMAR: An Open Language-Aware Multilingual Alignment Reranker}

\author{
    Seongtae Hong,
    Youngjoon Jang,
    Jungseob Lee,
    Seungyoon Lee,
    Heuiseok Lim\corresponding,
}
\affiliations{
    Korea University\\
    \texttt{\{ghdchlwls123, dew1701, omanma1928, dltmddbs100, limhseok\}@korea.ac.kr}
}

\begin{document}

\maketitle

\begin{abstract}
In multilingual retrieval augmented generation pipelines, an embedding model can retrieve relevant documents written in multiple languages, which are subsequently reranked before answer generation. However, it remains unclear whether existing multilingual rerankers consider document language when ordering semantically relevant candidates. Our analysis shows that these rerankers do not consistently prioritize documents written in the same language as the query when semantically equivalent documents are available across languages, even though document language can affect answer generation. We release LAMAR, a language aware multilingual cross encoder trained to account for both semantic relevance and language coherence. LAMAR first uses English anchored relevance distillation to establish consistent relevance scoring across multilingual inputs and then applies preference alignment for language coherence to encourage documents written in the same language as the query to receive higher rankings while retaining semantic relevance. In a controlled experiment designed to assess language coherence, LAMAR achieves the best performance overall and across all languages examined individually. LAMAR also remains competitive on established multilingual reranking benchmarks. In practical retrieval settings, LAMAR achieves the best results across all reported metrics when reranking candidates retrieved in the first stage. These results demonstrate that LAMAR accounts for language coherence while achieving strong performance on general multilingual reranking benchmarks.
\end{abstract}

\input{Texs/1.introduction}

\input{Texs/2.problem_analysis}

\input{Texs/3.method}

\input{Texs/4.Experimental_Settings}

\input{Texs/5.results}

\input{Texs/6.related_work}
\input{Texs/7.conclusion}


\bibliography{aaai2027}


\clearpage
\appendix

\input{appendix/appendix}

\end{document}

%% file: Texs/1.introduction.tex
\section{Introduction}
\label{sec:introduction}
Retrieval-Augmented Generation (RAG) mitigates the limitations of generation based solely on a large language model's internal knowledge by retrieving external documents and providing them as input to the model~\cite{rag1,rag2}. The answer generated by a RAG system depends strongly on whether the retrieved documents are relevant to the query~\cite{re2g,self-rag}. A first-stage retriever returns a set of candidate documents, not all of which are equally useful for generating the answer. A reranker therefore refines their ordering so that the most useful documents are placed at the top and provided to the LLM~\cite{rerank1,rerank2}. RAG systems are increasingly used in settings where user queries and knowledge sources span multiple languages~\cite{mrag1}. In multilingual RAG, the candidate set may contain semantically relevant documents written both in the query language and in other languages~\cite{xorqa,mrag2}. A multilingual reranker must therefore compare and order these documents across languages~\cite{qwen3}.

\begin{figure}[t]
    \centering
    \includegraphics[width=\columnwidth]{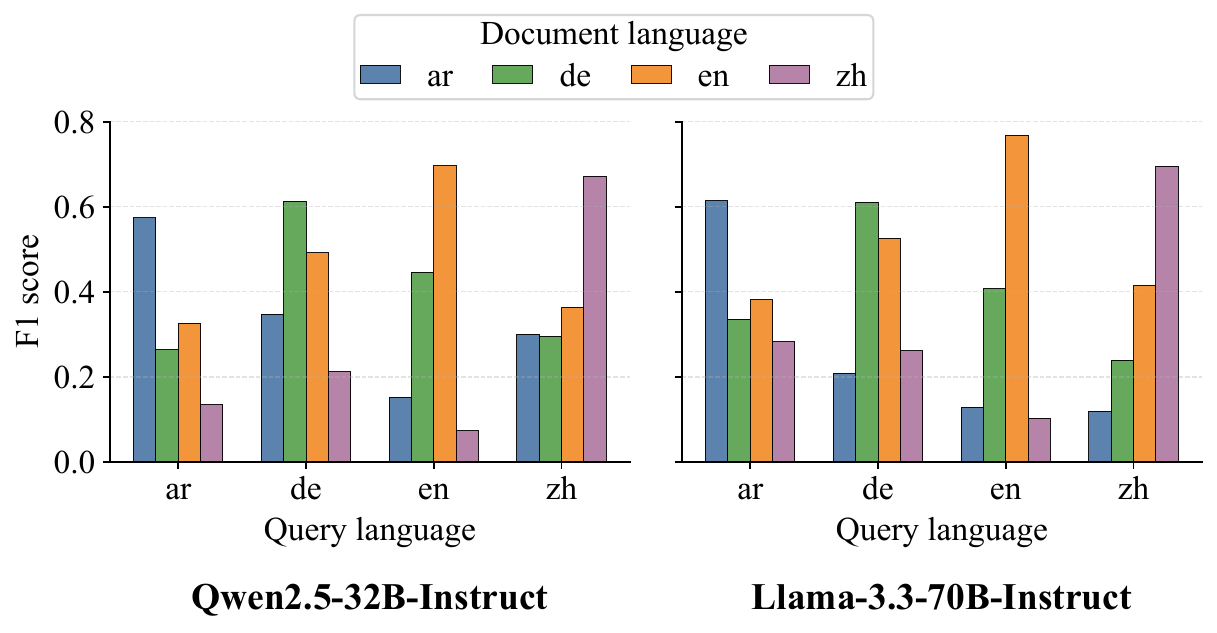}
  \caption{Multilingual RAG performance on XQuAD across query and document language combinations, measured by F1. Qwen2.5-32B-Instruct~\cite{qwen2.5} and Llama-3.3-70B-Instruct~\cite{llama3} generate answers from gold documents for all 16 combinations of Arabic, German, English, and Chinese.}
    \label{fig:intro_rag_f1}
\end{figure}

\begin{figure*}[t!]
\centering
\includegraphics[width=\textwidth]{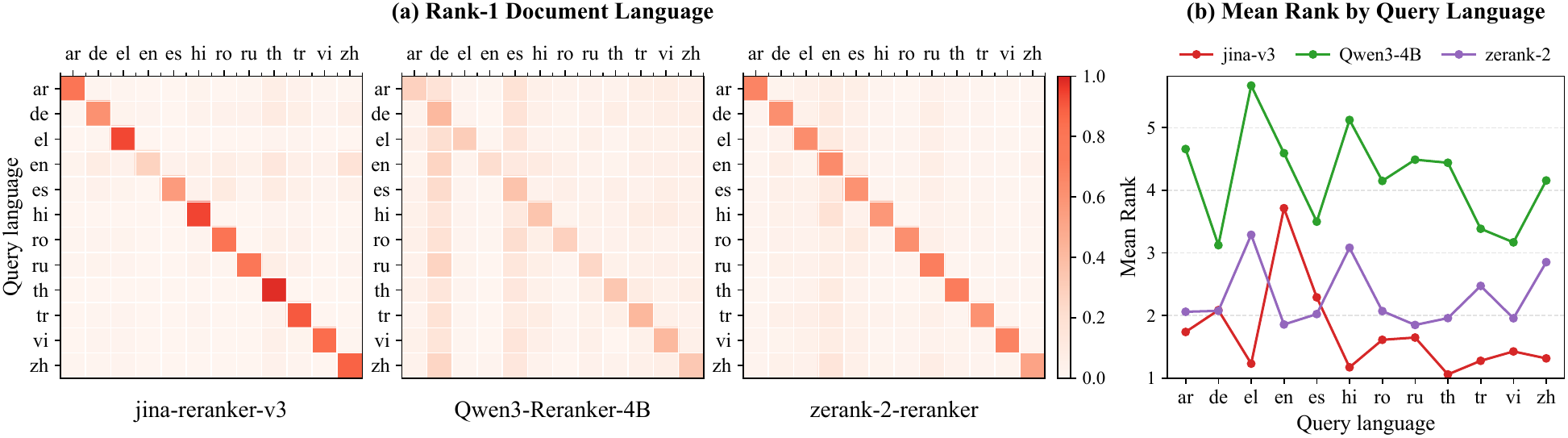}
\caption{Language coherence of Jina-reranker-v3~\cite{jinav3}, Qwen3-Reranker-4B~\cite{qwen3}, and zerank-2-reranker~\cite{zerank} across twelve languages using oracle candidate sets from XQuAD. (a) Top-1 document language distributions. Rows correspond to query languages, columns correspond to document languages, and each cell reports how frequently a document in the corresponding language is ranked first. (b) Mean rank of the gold document in the same language as the query. Stronger language coherence corresponds to diagonal concentration in (a) and lower mean ranks in (b).}

\label{fig:problem_analysis}
\end{figure*}

Although retrieving and ranking relevant documents are essential, the ultimate goal of multilingual RAG is to generate an accurate answer from those documents. Even when a retrieved document is semantically relevant, answer generation can vary depending on the language in which it is written~\cite{mrag1,pref1,pref2}. To examine this effect, we construct an evaluation from a multilingual parallel dataset by pairing each query with semantically equivalent gold documents written both in the query language and in other languages. As shown in Figure~\ref{fig:intro_rag_f1}, documents written in the same language as the query tend to yield higher F1 scores. Language consistency between the query and document is therefore an important consideration when determining which retrieved documents are provided to the generation model.

These findings raise the question of how a reranker should order semantically equivalent documents available in multiple languages. It remains unclear whether existing multilingual rerankers account for language coherence alongside semantic relevance, i.e., whether they identify the language of the query and prioritize documents written in that language. Our analysis shows that existing rerankers do not consistently place these documents at the top. This tendency is particularly pronounced for English queries, where documents written in other languages are often ranked above English documents. These results indicate that supporting multiple languages does not by itself ensure language coherence when training focuses primarily on semantic relevance.

In this paper, we release \textbf{LAMAR}, a reranker trained to jointly account for multilingual semantic relevance and language coherence. LAMAR is designed as a language-aware reranking model that can be used in multilingual RAG scenarios. To this end, LAMAR follows a two-stage training procedure. First, English-anchored relevance distillation calibrates semantic relevance scores across multilingual  input pairs. Second, preference alignment for language coherence combines a parallel document group ranking loss with a language coherence loss. Through these stages, LAMAR learns to rank relevant documents above non-relevant ones and, among semantically corresponding documents, prioritize those in the same language as the query.

Compared with 13 multilingual rerankers of varying model sizes, LAMAR more effectively accounts for the language of the query by prioritizing semantically equivalent documents written in that language on parallel multilingual oracle candidate sets. LAMAR also achieves competitive performance on established multilingual reranking benchmarks and exhibits consistent semantic reranking performance across languages. Finally, LAMAR records the highest scores across all reported metrics on candidate sets returned by first-stage retrievers, demonstrating robust language coherence and reranking effectiveness under practical retrieval conditions.

%% file: Texs/2.problem_analysis.tex
\section{Problem Analysis}
\label{sec:problem_analysis}
Multilingual rerankers are trained to compare queries and documents across languages and assign higher scores to semantically relevant pairs. Under this relevance-centered objective, documents in different languages can all be treated as valid candidates when they are semantically relevant to the query. In multilingual RAG, however, semantically equivalent documents may be available in multiple languages for the same query, and which language a document is written in can affect its suitability for answer generation. A reranker should therefore consider not only semantic relevance but also whether the query and document are in the same language. We refer to this consistency between the query and document languages as \emph{language coherence}. A language-coherent reranker should rank relevant documents above non-relevant ones and, among semantically equivalent documents, place those written in the query language closer to the top.

\paragraph{Diagnostic Evaluation Design.}
To examine language coherence as a reranker capability distinct from overall reranking effectiveness, we construct a diagnostic evaluation based on the XQuAD~\cite{xquad} dataset. It provides parallel queries and documents across twelve languages: Arabic, German, Greek, English, Spanish, Hindi, Romanian, Russian, Thai, Turkish, Vietnamese, and Chinese. For each query, we collect its gold documents in each language to form an oracle candidate set. We then rank the oracle candidate set using the same query expressed in each of the twelve languages. Only the document in the same language as the query is designated as relevant for this evaluation. We report the mean rank of this document across queries and its rank-1 selection rate. This evaluation design assesses whether a reranker recognizes the language of the query and identifies the document in the same language among the oracle candidates.

\paragraph{Does Language Coherence Differ Across Query Languages?}
Figure~\ref{fig:problem_analysis} (a) presents heatmaps showing, for each query language, which document languages appear at the top of the ranking and how frequently. All documents in each oracle candidate set are semantically equivalent gold documents, allowing the comparison to focus on the language of the document that each model ranks first. The results indicate that the evaluated rerankers do not consistently identify and prioritize the document written in the same language as the query. For example, Jina-reranker-v3 ranks the document in the query language first for 97.7\%, 94.0\%, and 93.4\% of Thai, Hindi, and Greek queries, respectively, but this rate decreases to 27.2\% for English queries. Among English queries, Chinese and Thai documents are instead ranked first in 16.4\% and 13.3\%, respectively. A similar limitation is observed for Qwen3-Reranker-4B, for which the rank-1 rate is 20.9\% for English queries and does not exceed 42.0\% for any query language. These results show that language coherence is inconsistent across languages.

\paragraph{Where Does the Document in the Same Language as the Query Rank?}
The Top-1 results in Figure~\ref{fig:problem_analysis} (a) reveal that same-language selection is far from uniform across query languages. Figure~\ref{fig:problem_analysis} (b) further examines where the same-language document appears in the full ranking by reporting its mean rank for each query language. The mean-rank results reveal substantial variation in the average position of the same-language document across query languages. For Jina-reranker-v3, the mean rank is below 2.0 for nine languages, with English as a notable exception at 3.71. For Qwen3-Reranker-4B, the mean rank exceeds 4.0 for eight of the twelve languages and reaches 5.67 for Greek and 5.12 for Hindi. These results demonstrate that ranking behavior varies across query languages, both in how often the document in the same language is ranked first and in its average position in the overall ranking.

\paragraph{Implication.}
These observations suggest that multilingual reranking requires more than language-agnostic semantic relevance. Figure~\ref{fig:intro_rag_f1} shows that consistency between the query and document languages is associated with higher QA performance, whereas Figure~\ref{fig:problem_analysis} shows that existing multilingual rerankers do not consistently prioritize the gold document in the query language. This motivates a reranker that ranks documents by relevance while prioritizing those in the same language as the query when semantically equivalent candidates are available.

%% file: Texs/3.method.tex
\section{LAMAR}
\label{sec:method}
LAMAR is a cross-encoder-based language-aware multilingual reranker that encodes a query--document pair to produce a scalar relevance score and is trained in two stages. The first stage, \emph{English-anchored relevance distillation}, calibrates semantic relevance scores across multilingual inputs using an English anchor, and the second stage, \emph{preference alignment for language coherence}, encourages higher rankings for documents written in the query language while preserving semantic relevance.

\subsection{English-Anchored Multilingual Relevance Distillation}
\label{subsec:semantic_distillation}
Reranker distillation typically feeds the same query--document pair $(q,d)$ to the teacher and student, using the teacher's relevance score as the regression target for the student. In the first stage, LAMAR distills the teacher score obtained from an English pair into a semantically corresponding multilingual pair so that query--document relations across languages are represented within a consistent semantic relevance space. 
Specifically, each training instance consists of a semantically corresponding English anchor pair $(q^{\mathrm{en}},d^{\mathrm{en}})$ and multilingual pair $(q^{\ell_q},d^{\ell_d})$, where $\ell_q$ and $\ell_d$ denote the languages of the query and document, respectively. Let $T$ denote the teacher ranker and $S_\theta$ the student reranker. The teacher assigns the English anchor pair a relevance score. For the corresponding multilingual pair, the student predicts a score, with the teacher's output serving as the regression target.
Accordingly, the student is optimized using the following mean squared error objective:
\begin{equation}
    \mathcal{L}_{\mathrm{distill}}
    =
    \left(
        S_\theta(q^{\ell_q}, d^{\ell_d})
        -
        T(q^{\mathrm{en}}, d^{\mathrm{en}})
    \right)^2
    \label{eq:distill_loss_kor}
\end{equation}
Minimizing this objective encourages the student score to approximate the teacher score obtained from the English anchor pair. Consequently, semantically corresponding multilingual pairs are scored on a consistent semantic relevance scale regardless of the input languages.

\subsection{Preference Alignment for Language Coherence}
\label{subsec:language_coherence_alignment}
The second stage aligns preferences among parallel candidates with language coherence. To this end, we combine a group ranking loss with a language-coherence loss in a joint objective:
\begin{equation}
    \mathcal{L}_{\mathrm{align}}
    =
    \mathcal{L}_{\mathrm{rank}}
    +
    \lambda
    \mathcal{L}_{\mathrm{LC}}
    \label{eq:alignment_loss}
\end{equation}
For a source-language query $q_{\mathrm{src}}$, we construct a parallel document group $\mathcal{D}=\{d^{+}_{\mathrm{src}},d^{+}_{\mathrm{tgt}}, d^{-}_{\mathrm{src}},d^{-}_{\mathrm{tgt}}\}$ and denote their scores by
$s^{+}_{\mathrm{src}}$, $s^{+}_{\mathrm{tgt}}$, $s^{-}_{\mathrm{src}}$, and $s^{-}_{\mathrm{tgt}}$, respectively. Here, $\mathrm{src}$ and $\mathrm{tgt}$ indicate the source and target languages, while $+$ and $-$ indicate positive and negative documents. This preference is expressed by the following score ordering:
\begin{equation}
    s^{+}_{\mathrm{src}}
    \gtrsim
    s^{+}_{\mathrm{tgt}}
    \quad > \quad
    s^{-}_{\mathrm{src}}
    \gtrsim
    s^{-}_{\mathrm{tgt}}
\label{eq:preference_order}
\end{equation}

\paragraph{Parallel-document group ranking loss.}
To preserve the ordering between positive and negative documents without imposing a relative ordering between semantically corresponding source and target documents, we employ Approx Discounted Rank MSE (ADR-MSE) as a listwise ranking objective~\cite{rank-distill}. Specifically, we assign a relevance label of 1 to positive documents and 0 to negative documents. Accordingly, for the score vector $\mathbf{s}=[s^{+}_{\mathrm{src}},s^{+}_{\mathrm{tgt}}, s^{-}_{\mathrm{src}},s^{-}_{\mathrm{tgt}}]$, we define the corresponding relevance label vector as $\mathbf{y}=[1,1,0,0]$. We index the documents in $\mathcal{D}$ according to their order in $\mathbf{s}$, where $s_i$ denotes the score of the $i$-th document. Let $r_i(\mathbf{y})$ denote the average rank of the $i$-th document induced by $\mathbf{y}$. The model-induced rank of the $i$-th document is approximated from the scores as follows:
\begin{equation}
  \hat{r}_i(\mathbf{s})
  =
  1+
  \sum_{\substack{j=1 \\ j \neq i}}^{|\mathcal{D}|}
  \sigma\!\left(s_j-s_i\right)
  \label{eq:approximate_rank}
\end{equation}
where $\sigma$ denotes the sigmoid, and the group ranking loss is
\begin{equation}
  \mathcal{L}_{\mathrm{rank}}
  =
  \frac{1}{|\mathcal{D}|}
  \sum_{i=1}^{|\mathcal{D}|}
  w_i
  \left(
      r_i(\mathbf{y})-\hat{r}_i(\mathbf{s})
  \right)^2
  \label{eq:rank_loss}
\end{equation}
The weight $w_i=1/\log_2(r_i(\mathbf{y})+1)$ assigns greater weight to documents appearing higher in the reference ranking. This loss ranks positive documents above negative documents without separately specifying the relative order between semantically equivalent parallel documents.

\paragraph{Language-coherence loss.}
The parallel-document group ranking loss preserves the relative order between positive and negative documents but does not capture a preference for documents written in the same language as the query when semantically corresponding documents are expressed in different languages. We therefore introduce the following language-coherence loss to model this preference:
\begin{equation}
  \mathcal{L}_{\mathrm{LC}}
  =
  \frac{1}{2}
  \left[
      \mathrm{softplus}
      \left(s^{+}_{\mathrm{tgt}}-s^{+}_{\mathrm{src}}\right)
      +
      \mathrm{softplus}
      \left(s^{-}_{\mathrm{tgt}}-s^{-}_{\mathrm{src}}\right)
  \right]
  \label{eq:language_coherence}
\end{equation}
Applying softplus to each score difference incorporates language preference into the relative ordering of semantically corresponding document pairs in both the positive and negative groups, without imposing a hard-margin constraint.



%% file: Texs/4.Experimental_Settings.tex
\section{Experimental Setup}
\label{sec:experimental_setup_kor}
This section describes the training data and implementation details for LAMAR, along with the baselines and evaluation setup.

\begin{table}[t!]
    \centering
    \small
    \renewcommand{\arraystretch}{1.}
    \resizebox{\columnwidth}{!}{%
    \begin{tabular}{llccl}
    \toprule
    Stage & Dataset & \# Lang. & \# Instances & Training Instance \\
    \midrule
    \multirow{3}{*}{Stage 1}
        & MMARCO & 14 & 1.6M & \multirow{3}{*}{$\big((q_i^{\mathrm{en}},d_j^{\mathrm{en}}),(q_i^{\ell_q},d_j^{\ell_d})\big)$} \\
        & MIRACL & 51 & 343K & \\
        & RLHN & 1 & 4.6M & \\
    \midrule
    Stage 2 & MIRACL & 51 & 8.5K & $(q_{\mathrm{src}},d_{\mathrm{src}}^{+},d_{\mathrm{tgt}}^{+},d_{\mathrm{src}}^{-},d_{\mathrm{tgt}}^{-})$ \\
    \bottomrule
    \end{tabular}
    }
    \caption{Training data used for LAMAR, with language coverage, number of constructed instances, and input format reported for each dataset and training stage.}
    \label{tab:training_datasets}
\end{table}

\subsection{Training}
\paragraph{Dataset.}

We use MMARCO~\cite{mmarco}, multilingual triplets from a translated version of the MIRACL dataset~\cite{miracl}, and RLHN~\cite{rlhn} as training datasets. MMARCO and MIRACL provide semantically corresponding parallel data across multiple languages; MMARCO associates each query with one positive document, whereas MIRACL additionally provides one negative document. RLHN contains only English examples, with each query associated with multiple positive and negative documents. Based on these original structures, we reorganize each dataset into the language combinations and instance formats required by the corresponding training stage. Table~\ref{tab:training_datasets} summarizes the datasets used in each stage, the resulting training set sizes, and the corresponding instance formats. The final training sets contain 6.7M and 8.6K instances in Stages 1 and 2, respectively. In Stage 1, each instance consists of an English teacher pair and its corresponding multilingual student pair. For MMARCO and MIRACL, we use both the originally associated document and documents sampled from other rows of the same dataset; the teacher and student receive the English and multilingual realizations, respectively, of the same selected document. The query and document language combinations are balanced across the complete training set. For RLHN, each query is paired separately with every positive and negative document in its set, and the same English pair is provided to both the teacher and student. In Stage 2, each listwise instance is constructed from a single MIRACL triplet. Each triplet is used repeatedly while maintaining a balanced distribution of source and target language combinations across the training set. Dataset language coverage and the data synthesis algorithms used in each stage are provided in Appendix~A.

\setcounter{table}{1}
\begin{table}[t]
  \centering
  \renewcommand{\arraystretch}{1.}
  \resizebox{\columnwidth}{!}{%
  \begin{tabular}{lcl}
  \toprule
  Benchmark & \# Lang. & Languages \\
  \midrule
  \multicolumn{3}{c}{\textbf{\text{Language-Coherence Evaluation}}} \\
  XQuAD & 12 & ar, de, el, en, es, hi, ro, ru, th, tr, vi, zh \\
  BELEBELE & 14 & ar, de, en, es, fr, hi, id, it, ja, nl, pt, ru, vi, zh \\
  \midrule
  \multicolumn{3}{c}{\textbf{\text{Multilingual Reranking Evaluation}}} \\
  MIRACL & 18 & ar, bn, de, en, es, fa, fi, fr, hi, id, ja, ko, ru, sw, te, th, yo, zh \\
  XGLUE & 7 & de, en, es, fr, it, pt, zh \\
  HUME & 3 & da, en, no \\
  MLDR & 13 & ar, de, en, es, fr, hi, it, ja, ko, pt, ru, th, zh \\
  Wikipedia & 16 & bg, bn, cs, da, de, en, fa, fi, hi, it, nl, no, pt, ro, sr, sv \\
  \bottomrule
  \end{tabular}%
  }
  \caption{Benchmarks and language coverage for language-coherence and multilingual reranking evaluation.}
  \label{tab:evaluation_benchmarks}
\end{table}

\subsection{Training Details}
LAMAR is initialized from the multilingual base encoder bge-m3-retromae~\cite{bge}, and all training is conducted on eight NVIDIA RTX A6000 GPUs. Both stages are trained for one epoch with a maximum sequence length of 8,192, a warmup ratio of 0.1, the AdamW optimizer, a linear decay scheduler, and bf16 precision. In Stage 1, Qwen3-Reranker-4B~\cite{qwen3} is used as the teacher, with a learning rate of $2\times10^{-5}$ and a batch size of 384. In Stage 2, the coefficient $\lambda$ for $\mathcal{L}_{\mathrm{LC}}$ is set to 2, with a learning rate of $1\times10^{-6}$ and a batch size of 32.

\setcounter{table}{2}
\input{Tables/Main}

\subsection{Evaluation}
\label{subsec:evaluation}

\paragraph{Baselines.}
We compare LAMAR with the following publicly available multilingual rerankers: bge-reranker-v2-m3/gemma~\cite{bge}, jina-reranker-v2-base-multilingual, jina-reranker-v3~\cite{jinav3}, gte-multilingual-reranker-base~\cite{gte}, Qwen3-Reranker-0.6B/4B~\cite{qwen3}, llama-nemotron-rerank-1b-v2, ctxl-rerank-v2-instruct-multilingual-1b~\cite{ctxl}, zerank-2-reranker~\cite{zerank}, and Prism-Qwen3.5-Reranker-0.8B/2B/4B~\cite{prism}.

\begin{figure*}[th]
    \centering
    \includegraphics[width=\textwidth]{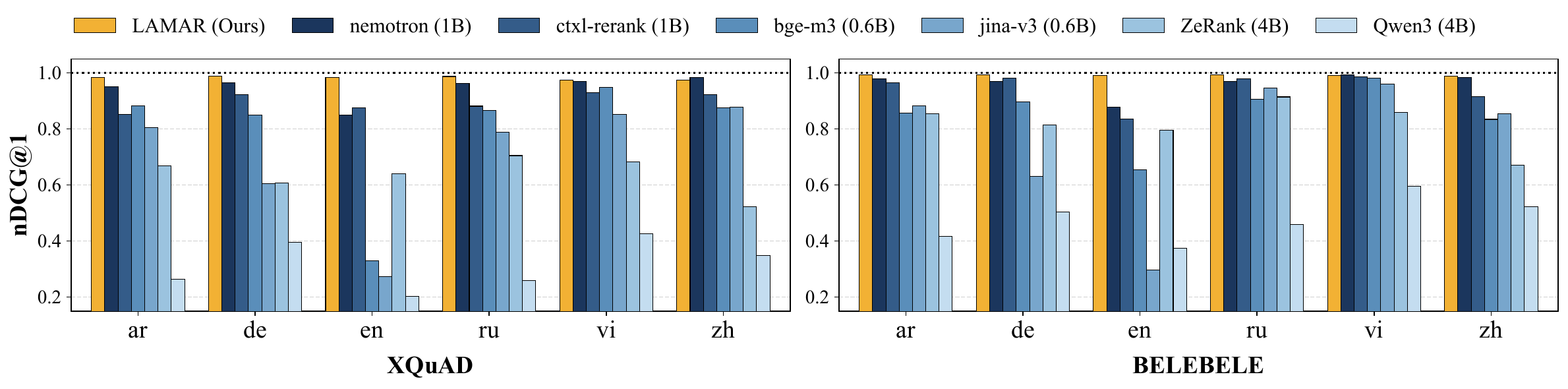}
    \caption{nDCG@1 for the six query languages common to XQuAD and BELEBELE under the parallel oracle setting. While results are reported for the shared query languages, evaluation retains the oracle candidate sets used in the overall results, containing documents in 12 and 14 languages for XQuAD and BELEBELE, respectively.}
    \label{fig:lang}
\end{figure*}

\paragraph{Benchmarks.}
We conduct evaluation in two settings. Table~\ref{tab:evaluation_benchmarks} summarizes the benchmarks and evaluation languages included in each setting. (1) The oracle reranking evaluation measures language aware behavior. Following the setup used in the problem analysis, we use oracle subsets constructed from two multilingual benchmarks, XQuAD~\cite{xquad} and BELEBELE~\cite{belebele}, both of which provide parallel queries and documents in multiple languages. For each query, the candidate set consists exclusively of semantically equivalent documents that are correct for the query and expressed in different languages. We designate the document written in the same language as the query as relevant and use nDCG@1, nDCG@10, and MRR@10 to measure how highly it is ranked. (2) The standard multilingual reranking evaluation measures reranking performance across multiple languages using MIRACL~\cite{miracl}, XGLUE~\cite{xglue}, HUME~\cite{hume2}, MLDR~\cite{bge}, and Wikipedia~\cite{wikipedia} from the MMTEB~\cite{mmteb} multilingual reranking benchmarks. We report nDCG@10 as the primary metric. These complementary settings assess the robustness of LAMAR across distinct multilingual reranking scenarios.

%% file: Tables/Main.tex
\begin{table*}[t!]
\centering
\renewcommand{\arraystretch}{1.0}
\resizebox{\textwidth}{!}{
\begin{tabular}{lccccccc|ccc}
\toprule
\multirow{2}{*}{\textbf{Model}} & \multirow{2}{*}{\textbf{Size}} 
& \multicolumn{3}{c}{\textbf{XQuAD (12)}} 
& \multicolumn{3}{c|}{\textbf{BELEBELE (14)}}
& \multicolumn{2}{c}{\textbf{nDCG Avg.}} 
& \multirow{2}{*}{\textbf{Rank}} \\
\cmidrule(lr){3-5} \cmidrule(lr){6-8} \cmidrule(lr){9-10}
& & \textbf{nDCG@1} & \textbf{nDCG@10} & \textbf{MRR@10}
& \textbf{nDCG@1} & \textbf{nDCG@10} & \textbf{MRR@10} 
& \textbf{@1} & \textbf{@10} & \\
\midrule
gte-multilingual-reranker-base & 0.3B & 82.59 & 91.51 & 88.89 & 77.05 & 88.58 & 85.08 & 79.82 & 90.05 & 6 \\
jina-reranker-v2-base-multilingual & 0.3B & 57.30 & 79.36 & 72.55 & 60.47 & 79.88 & 73.49 & 58.89 & 79.62 & 13 \\
bge-reranker-v2-m3 & 0.6B & 84.67 & 92.20 & 89.89 & 84.83 & 92.65 & 90.32 & 84.75 & 92.43 & 5 \\
Qwen3-Reranker-0.6B & 0.6B & 61.72 & 79.58 & 73.56 & 67.67 & 82.85 & 77.78 & 64.70 & 81.22 & 12 \\
jina-reranker-v3 & 0.6B & 77.75 & 88.31 & 84.79 & 78.10 & 88.43 & 85.10 & 77.93 & 88.37 & 7 \\
Prism-Qwen3.5-Reranker-0.8B & 0.8B & 72.28 & 86.74 & 82.21 & 71.11 & 85.81 & 79.70 & 71.70 & 86.28 & 8 \\
llama-nemotron-rerank-1b-v2 & 1B & \underline{95.83} & \underline{98.02} & \underline{97.39} & 94.32 & 97.21 & 96.37 & \underline{95.08} & \underline{97.62} & 2 \\
ctxl-rerank-v2-instruct-multilingual-1b & 1B & 88.79 & 94.92 & 93.27 & \underline{94.64} & \underline{97.54} & \underline{96.72} & 91.72 & 96.23 & 3 \\
Prism-Qwen3.5-Reranker-2B & 2B & 70.41 & 85.34 & 79.58 & 60.62 & 79.62 & 70.27 & 65.52 & 82.48 & 11 \\
bge-reranker-v2-gemma & 2B & 86.63 & 93.42 & 91.24 & 84.55 & 92.49 & 90.04 & 85.59 & 92.96 & 4 \\
Qwen3-Reranker-4B & 4B & 33.03 & 58.81 & 48.61 & 43.31 & 67.02 & 58.79 & 38.17 & 62.92 & 14 \\
zerank-2-reranker & 4B & 63.13 & 79.98 & 74.31 & 74.71 & 87.16 & 83.22 & 68.92 & 83.57 & 10 \\
Prism-Qwen3.5-Reranker-4B & 4B & 70.53 & 85.19 & 80.64 & 67.32 & 83.23 & 76.95 & 68.93 & 84.21 & 9 \\
\hline
\textbf{LAMAR (Ours)} & \textbf{0.6B} & \textbf{96.89} & \textbf{98.59} & \textbf{98.10} & \textbf{94.66} & \textbf{97.60} & \textbf{96.79} & \textbf{95.78} & \textbf{98.10} & \textbf{1} \\
\bottomrule
\end{tabular}
}
\caption{Language-coherence results on the parallel oracle subsets of XQuAD and BELEBELE. The nDCG averages are computed over XQuAD and BELEBELE, and Rank is determined by averaging the @1 and @10 scores. Best and second-best results are shown in bold and underlined, respectively.}

\label{tab:main_xquad_belebele}
\end{table*}

%% file: Texs/5.results.tex
\input{Tables/miracl}

\input{Tables/multi_rank}

\section{Experimental Results}
\label{sec:results_kor}

\subsection{Language-Coherence Evaluation}
\label{subsec:results_oracle_kor}

\paragraph{Overall Performance.}
Table~\ref{tab:main_xquad_belebele} presents results on the parallel oracle subsets of XQuAD and BELEBELE, with the document written in the same language as the query designated as relevant for evaluation. Because these subsets consist exclusively of passages that convey semantically equivalent answer content across multiple languages, they enable direct evaluation of language coherence while controlling for semantic relevance. LAMAR achieves the best performance on both benchmarks. It records nDCG@1 and @10 scores of 0.9689 and 0.9859 on XQuAD, respectively, and 0.9466 and 0.9760 on BELEBELE. These results show that LAMAR is the most effective at placing the document written in the same language as the query at the top among semantically equivalent multilingual candidates. 

Notably, the differences between models are substantially larger at nDCG@1 than at nDCG@10. Because all candidate documents in the parallel oracle subsets are semantically equivalent, variation in the top-ranked document is more directly associated with which document language a model prioritizes than with ordinary relevance discrimination. Consistent with our problem analysis, these results demonstrate that rerankers trained primarily to model semantic relevance do not necessarily exhibit language coherence and that LAMAR reflects language coherence more consistently than existing rerankers.

\paragraph{Consistency Across Languages.}
Figure~\ref{fig:lang} compares nDCG@1 for each of the six languages common to both benchmarks. LAMAR records the highest nDCG@1 in all six languages on both benchmarks, while its performance remains stable across languages. Other rerankers generally achieve lower scores and exhibit substantial performance gaps between languages, particularly for English queries. For example, despite achieving high scores in the other five languages, llama-nemotron-rerank-1b-v2 records substantially lower English nDCG@1 scores of 0.8496 on XQuAD and 0.8789 on BELEBELE. Overall, these results demonstrate that LAMAR provides robust language-coherence performance across diverse query languages. Results for all languages and metrics are provided in Appendix~B..

\subsection{Multilingual Reranking Performance}
Table~\ref{tab:mteb_multilingual_reranking} presents results on five MTEB multilingual reranking benchmarks. LAMAR achieves an average score of 86.84, the second-highest overall result. Its performance remains competitive across benchmarks with different datasets and language coverage. Notably, LAMAR has 0.6B parameters but performs comparably to substantially larger rerankers, showing that its language-aware behavior is accompanied by strong general semantic reranking effectiveness.

We further analyze MIRACL to examine semantic reranking performance within individual languages. MIRACL follows a multi-monolingual setting in which relevance is evaluated within the query and document collection of each of its 18 languages. Table~\ref{tab:miracl_reranker_by_size} shows that LAMAR records an average score of 69.5, the second-highest average in the comparison. LAMAR also remains competitive across linguistically diverse languages, including English, Finnish, Telugu, and Thai, rather than exhibiting strong performance only for a particular language group. Overall, the results demonstrate that LAMAR combines language coherence with competitive multilingual semantic reranking performance.

\input{Tables/subset}

\subsection{Robustness under Retrieved Candidate Sets}
\label{subsec:results_subset_kor}
The preceding experiments use controlled oracle subsets containing candidates from all languages, enabling direct analysis of language coherence. In practical retrieval pipelines, reranking is performed over candidate sets returned by a retriever. We therefore comprehensively evaluate LAMAR in a setting designed to assess both reranking effectiveness and language coherence. Table~\ref{tab:xquad_retriever_subset} compares XQuAD reranking performance on the top-20 candidate sets returned by multilingual-MiniLM-L12-v2 and bge-m3 for queries in each language. LAMAR achieves the best performance under all three metrics on the candidate subsets from both retrievers, demonstrating consistent language coherence and reranking performance even when the candidate composition changes.

\begin{figure}[t!]
    \centering
    \includegraphics[width=\columnwidth]{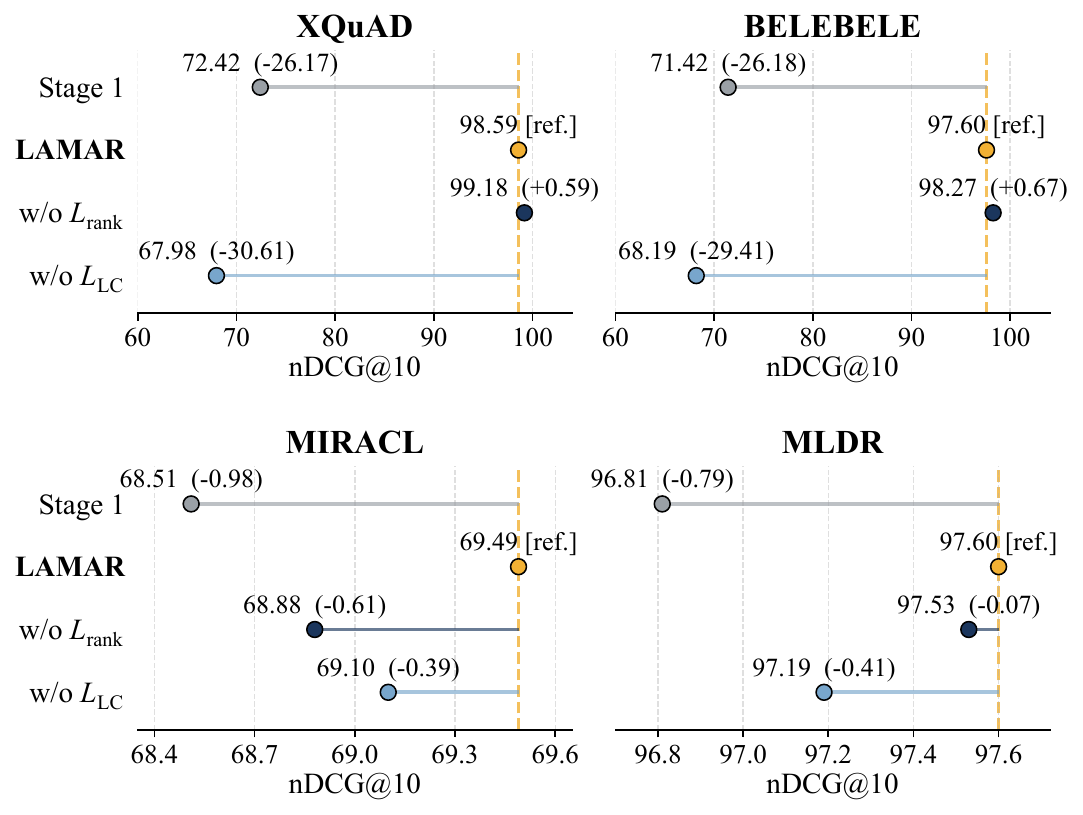}
    \caption{nDCG@10 ablation results across LAMAR training stages and Stage 2 loss configurations. Dashed lines denote the full LAMAR model (Stage 1+2), with absolute differences shown in parentheses.}
    \label{fig:ablation_lamar}
\end{figure}

\subsection{Ablation Study}
\label{subsec:results_ablation_kor}
Figure~\ref{fig:ablation_lamar} analyzes how the performance of the final model varies across training stages and objective components. Through English-anchored relevance distillation, the Stage 1 model learns a consistent semantic relevance scale across multilingual inputs and achieves competitive reranking performance, but its ability to reflect language preference in document ordering remains limited in the absence of an explicit preference objective. LAMAR, trained with both stages, preserves semantic reranking performance while substantially improving language-aware ordering.

The two ablation variants reveal the contribution of each loss in the Stage 2 objective. Removing $\mathcal{L}_{\mathrm{LC}}$ substantially degrades performance on language-coherence evaluation, indicating that this term is essential for learning a preference for documents written in the same language as the query. Conversely, the variant without $\mathcal{L}_{\mathrm{rank}}$ outperforms the final model on language-coherence evaluation. The general multilingual reranking results, however, suggest that $\mathcal{L}_{\mathrm{rank}}$ plays an important role in assigning higher relevance scores to positive documents than to negative documents. Together, the two components enable LAMAR to balance language coherence with general multilingual reranking performance.

%% file: Tables/miracl.tex
\begin{table*}[t!]
\centering
\renewcommand{\arraystretch}{1}
\resizebox{\textwidth}{!}{
\begin{tabular}{lccc|cccccccccccccccccc}
\toprule
\textbf{Model} & \textbf{Size} & \textbf{Avg.} & \textbf{Rank} & \textbf{ar} & \textbf{bn} & \textbf{de} & \textbf{en} & \textbf{es} & \textbf{fa} & \textbf{fi} & \textbf{fr} & \textbf{hi} & \textbf{id} & \textbf{ja} & \textbf{ko} & \textbf{ru} & \textbf{sw} & \textbf{te} & \textbf{th} & \textbf{yo} & \textbf{zh} \\
\midrule
gte-multi-reranker-base & 0.3B & 67.3 & 8 & 77.8 & 78.1 & 53.2 & 66.4 & 63.3 & 58.6 & 80.2 & 56.1 & 65.0 & 62.2 & 70.3 & 69.8 & 66.3 & 64.7 & 79.8 & 77.3 & 68.2 & 54.9 \\
jina-reranker-v2-base-multi & 0.3B & 67.9 & 7 & 76.9 & 80.0 & 54.6 & 65.4 & 64.8 & 59.8 & 79.1 & 55.7 & 64.2 & 63.3 & 70.5 & 72.7 & 65.1 & 67.9 & 82.6 & 79.1 & 66.0 & 53.9 \\
bge-reranker-v2-m3 & 0.6B & 69.1 & 4 & 79.2 & 80.2 & 56.3 & 66.0 & 67.3 & 61.7 & 81.2 & 58.7 & 68.3 & 65.4 & 71.0 & 69.9 & 67.4 & 69.8 & 79.5 & 78.7 & 69.9 & 54.2 \\
Qwen3-Reranker-0.6B & 0.6B & 64.9 & 10 & 76.5 & 65.0 & 55.5 & 65.6 & 66.9 & 60.5 & 79.2 & 55.7 & 62.0 & 62.8 & 69.2 & 72.8 & 64.2 & 60.9 & 52.8 & 77.8 & 65.4 & 55.3 \\
jina-reranker-v3 & 0.6B & 68.6 & 6 & 80.0 & 79.0 & 57.0 & 67.8 & 66.4 & 59.2 & 80.0 & 56.3 & 63.3 & 64.8 & 72.3 & 73.3 & 67.9 & 65.6 & 81.2 & 79.8 & 64.8 & 55.7 \\
Prism-Qwen3.5-Reranker-0.8B & 0.8B & 52.1 & 14 & 67.8 & 35.1 & 47.3 & 60.3 & 54.9 & 48.5 & 71.5 & 48.1 & 18.9 & 53.8 & 63.7 & 63.8 & 59.7 & 52.3 & 32.1 & 52.1 & 57.4 & 51.0 \\
llama-nemotron-rerank-1b-v2 & 1B & 69.2 & 3 & 80.7 & 77.7 & 57.0 & 66.6 & 67.9 & 61.2 & 82.5 & 57.9 & 67.2 & 63.6 & 72.7 & 70.9 & 68.8 & 68.0 & 79.6 & 80.4 & 67.7 & 55.9 \\
ctxl-rerank-v2-inst-multi-1b & 1B & 65.8 & 9 & 79.8 & 69.1 & 57.1 & 68.1 & 64.8 & 55.9 & 81.1 & 57.6 & 55.8 & 64.3 & 72.5 & 71.3 & 68.8 & 62.5 & 57.4 & 78.8 & 65.0 & 54.8 \\
Prism-Qwen3.5-Reranker-2B & 2B & 57.0 & 13 & 70.6 & 54.1 & 52.9 & 61.8 & 59.7 & 53.8 & 75.3 & 51.8 & 34.0 & 55.5 & 67.1 & 65.8 & 61.5 & 57.4 & 45.0 & 52.7 & 54.7 & 51.8 \\
bge-reranker-v2-gemma & 2B & \textbf{69.7} & \textbf{1} & 80.9 & 80.6 & 57.2 & 67.3 & 67.7 & 61.8 & 82.1 & 59.0 & 68.0 & 62.9 & 73.6 & 71.8 & 68.8 & 68.8 & 83.6 & 80.8 & 65.9 & 54.4 \\
Qwen3-Reranker-4B & 4B & 68.9 & 5 & 81.0 & 70.7 & 59.5 & 70.3 & 67.2 & 64.2 & 83.9 & 62.0 & 65.1 & 64.2 & 76.2 & 73.7 & 69.4 & 68.2 & 58.3 & 81.1 & 68.3 & 56.6 \\
Prism-Qwen3.5-Reranker-4B & 4B & 58.0 & 12 & 71.3 & 51.4 & 54.3 & 62.4 & 58.4 & 50.7 & 77.3 & 54.4 & 32.6 & 55.7 & 68.4 & 66.8 & 64.2 & 62.3 & 48.5 & 55.0 & 59.8 & 51.4 \\
zerank-2-reranker & 4B & 63.3 & 11 & 76.7 & 68.3 & 55.0 & 64.2 & 61.5 & 54.8 & 79.6 & 55.4 & 56.6 & 56.7 & 70.7 & 67.0 & 66.7 & 58.7 & 53.7 & 76.7 & 63.4 & 53.6 \\
\hline
\textbf{LAMAR (Ours)} & 0.6B & \underline{69.5} & \underline{2} & 78.9 & 79.3 & 56.6 & 68.8 & 66.2 & 62.5 & 82.3 & 59.9 & 67.8 & 62.9 & 71.1 & 71.1 & 67.5 & 68.7 & 83.1 & 80.4 & 68.4 & 55.4 \\
\bottomrule
\end{tabular}
}
\caption{nDCG@10 results on MIRACL across 18 languages under the multi-monolingual setting. Rank is determined by the average nDCG@10 across all languages. Best and second-best results are shown in bold and underlined, respectively.}
\label{tab:miracl_reranker_by_size}
\end{table*}

%% file: Tables/multi_rank.tex
\begin{table}[t]
\centering
\renewcommand{\arraystretch}{1.}
\resizebox{\columnwidth}{!}{
\begin{tabular}{l *{5}{>{\centering\arraybackslash}p{0.95cm}}|c}
\toprule
\textbf{Model} & \textbf{Miracl} & \textbf{Xglue} & \textbf{Hume} & \textbf{Mldr} & \textbf{Wiki} & \textbf{Avg.} \\
\midrule
gte-multi-reranker-base & 67.34 & 76.27 & 94.23 & 98.81 & 92.11 & 85.75 \\
jina-reranker-v2-base-multi & 67.86 & 76.77 & 93.41 & 86.17 & 93.96 & 83.63 \\
bge-reranker-v2-m3 & 69.15 & 76.49 & 94.55 & 97.40 & 93.96 & 86.31 \\
Qwen3-Reranker-0.6B & 64.88 & 76.52 & 94.60 & \underline{98.85} & 94.69 & 85.91 \\
jina-reranker-v3 & 68.56 & \textbf{80.69} & 94.79 & 93.41 & \underline{94.83} & 86.46 \\
Prism-Qwen3.5-Reranker-0.8B & 52.11 & 74.11 & 94.11 & 96.35 & 92.80 & 81.90 \\
llama-nemotron-rerank-1b-v2 & 69.23 & 77.95 & \underline{96.02} & 96.42 & 94.50 & 86.82 \\
ctxl-rerank-v2-inst-multi-1b & 65.80 & 77.81 & 93.60 & 97.37 & 93.65 & 85.65 \\
Prism-Qwen3.5-Reranker-2B & 56.96 & 74.80 & 95.15 & 96.46 & 93.79 & 83.43 \\
bge-reranker-v2-gemma & \textbf{69.72} & 76.75 & 94.21 & 89.07 & 94.63 & 84.88 \\
Qwen3-Reranker-4B & 68.88 & 76.40 & 95.73 & \textbf{99.39} & \textbf{96.31} & \textbf{87.34} \\
Prism-Qwen3.5-Reranker-4B & 58.03 & 74.96 & 94.78 & 97.32 & 94.28 & 83.87 \\
zerank-2-reranker & 63.31 & \underline{78.01} & \textbf{96.13} & 98.03 & 94.51 & 86.00 \\
\hline
\textbf{LAMAR (Ours)} & \underline{69.49} & 77.03 & 95.30 & 97.60 & 94.79 & \underline{86.84} \\
\bottomrule
\end{tabular}
}
\caption{Performance comparison on five MTEB multilingual reranking benchmarks using nDCG@10.}
\label{tab:mteb_multilingual_reranking}
\end{table}

%% file: Tables/subset.tex
\begin{table}[t]
\centering
\renewcommand{\arraystretch}{1.15}
\resizebox{\columnwidth}{!}{
\begin{tabular}{l ccc ccc}
\toprule
\multirow{3}{*}{\textbf{Model}}
& \multicolumn{3}{c}{\textbf{M-MiniLM-v2}}
& \multicolumn{3}{c}{\textbf{bge-m3}} \\
& \multicolumn{3}{c}{(R@20: 80.67)}
& \multicolumn{3}{c}{(R@20: 97.30)} \\
\cmidrule(lr){2-4} \cmidrule(lr){5-7}
& \textbf{N@1} & \textbf{N@10} & \textbf{M@10}
& \textbf{N@1} & \textbf{N@10} & \textbf{M@10} \\
\midrule
gte-multi-reranker-base
& 67.35 & 74.26 & 71.96
& 79.77 & 88.76 & 86.12 \\

jina-reranker-v2-base-multi
& 50.68 & 66.50 & 61.73
& 56.09 & 77.40 & 71.03 \\

bge-reranker-v2-m3
& 69.85 & 75.23 & 73.18
& 83.27 & 90.20 & 88.10 \\

Qwen3-Reranker-0.6B
& 50.81 & 64.89 & 60.17
& 59.34 & 76.78 & 71.08 \\

jina-reranker-v3
& 61.74 & 70.98 & 68.02
& 68.40 & 82.04 & 77.54 \\

Prism-Qwen3.5-Reranker-0.8B
& 60.48 & 71.35 & 66.64
& 68.81 & 83.91 & 79.05 \\

llama-nemotron-rerank-1b-v2
& \underline{77.32} & \underline{79.17} & \underline{78.66}
& \underline{92.48} & \underline{95.10} & \underline{94.39} \\

ctxl-rerank-v2-inst-multi-1b
& 69.56 & 75.73 & 74.09
& 80.90 & 89.95 & 87.54 \\

Prism-Qwen3.5-Reranker-2B
& 60.46 & 70.93 & 65.75
& 70.08 & 84.00 & 79.31 \\

bge-reranker-v2-gemma
& 70.94 & 76.04 & 74.45
& 83.90 & 90.79 & 88.65 \\

Qwen3-Reranker-4B
& 29.09 & 50.29 & 42.02
& 33.61 & 58.73 & 49.19 \\

zerank-2-reranker
& 52.02 & 65.46 & 61.04
& 60.98 & 77.55 & 72.09 \\

Prism-Qwen3.5-Reranker-4B
& 60.89 & 71.06 & 66.08
& 70.25 & 84.01 & 79.32 \\

\hline
\textbf{LAMAR (Ours)}
& \textbf{77.91} & \textbf{79.46} & \textbf{79.06}
& \textbf{92.90} & \textbf{95.36} & \textbf{94.72} \\
\bottomrule
\end{tabular}
}
\caption{XQuAD reranking performance on top-20 candidate subsets retrieved by paraphrase-multilingual-MiniLM-L12-v2 (M-MiniLM-v2) and bge-m3~\cite{bge}. R, N, and M denote Recall, nDCG, and MRR, respectively.}
\label{tab:xquad_retriever_subset}
\end{table}

%% file: Texs/6.related_work.tex
\section{Related Work}
\label{sec:related_work}

\subsection{Language Awareness in Multilingual RAG}
Multilingual RAG systems must retrieve information and generate answers across languages~\cite{xrag,mrag1}. Prior work has shown that retrievers may favor particular document languages even when candidates provide comparable semantic information~\cite{investigate}, while the language of the retrieved context can affect generation quality and evidence use~\cite{pref1,m-consistency}. These findings have motivated language-aware evaluation protocols~\cite{malire}, representation alignment through distillation or cross-lingual training~\cite{distill,cross-hong}, and translation, evidence fusion, and query augmentation for multilingual RAG~\cite{investigate,m-debiased}. Recent work has studied language bias in reranking~\cite{pref2}, but the broader literature has primarily focused on first-stage retrieval or answer generation, leaving the reranking stage comparatively underexplored.

\subsection{Multilingual Rerankers}
Prior work on multilingual reranking has explored parameter efficient transfer~\cite{transfer}, zero shot reranking with large language models~\cite{rankgpt}, translated supervision and listwise training~\cite{m-reranker}, and systematic  evaluation of pairwise and listwise rerankers~\cite{m-reranker2}. Recent models have further advanced multilingual reranking through multilingual  pretraining, long context modeling, and supervision from large language models~\cite{gte,jinav3,qwen3}. Knowledge distillation provides a complementary direction by transferring ranking signals from stronger teachers to efficient student models~\cite{scoredistill,rank-distill}, including the transfer of English ranking knowledge to multilingual inputs~\cite{cross-distill}. Despite these advances, existing approaches primarily focus on relevance estimation across languages, without explicitly determining how semantically corresponding documents expressed in different languages should be ordered. LAMAR addresses this gap through English anchored relevance distillation and preference alignment for language coherence.

%% file: Texs/7.conclusion.tex
\section{Conclusion}
\label{sec:conclusion}
This work examines whether multilingual rerankers account for language coherence when semantically equivalent documents are available in multiple languages. Our analysis shows that existing rerankers do not consistently prioritize documents written in the same language as the query, while the multilingual rag evaluation further indicates that document language can affect answer generation. We introduce LAMAR, a language aware multilingual cross encoder trained through English anchored relevance distillation and language coherence preference alignment. LAMAR achieves the highest overall nDCG@1 in the controlled language coherence evaluation, remains competitive on general multilingual reranking benchmarks, and obtains the best results under retrieved candidate settings. The ablation study further confirms the complementary contributions of the relevance ranking and language coherence objectives. Overall, our findings establish language coherence as a key capability for multilingual rerankers and highlight the importance of considering it alongside semantic relevance when ranking documents for multilingual RAG.

%% file: appendix/appendix.tex
\section{Training Dataset Details}
This section details the sources and language coverage of the datasets used to train LAMAR, together with the instance synthesis procedures for its two training stages. After presenting the language coverage and public links for each dataset, we describe the data construction algorithms and dataset-specific synthesis settings used for English-anchored relevance distillation and preference alignment for language coherence.

\begin{table}[H]
    \centering
    \small
    \setlength{\tabcolsep}{3pt}
    \begin{tabular}{@{}l c >{\raggedright\arraybackslash}p{0.65\columnwidth}@{}}
    \toprule
    Dataset & \# Lang. & Languages \\
    \midrule
    RLHN & 1 & English. \\
    \addlinespace
    MMARCO & 14 & Arabic, Chinese, Dutch, English, French, German, Hindi, Indonesian, Italian, Japanese, Portuguese, Russian, Spanish, and Vietnamese. \\
    \addlinespace
    MIRACL & 51 & Arabic, Bengali, Bulgarian, Catalan, Chinese, Croatian, Czech, Danish, Dutch, English, Estonian, Filipino, Finnish, French, German, Greek, Gujarati, Hebrew, Hindi, Hungarian, Icelandic, Indonesian, Italian, Japanese, Kannada, Korean, Latvian, Lithuanian, Malayalam, Marathi, Norwegian, Persian, Polish, Portuguese, Punjabi, Romanian, Russian, Serbian, Slovak, Slovenian, Spanish, Swahili, Swedish, Tamil, Telugu, Thai, Turkish, Ukrainian, Urdu, Vietnamese, and Zulu. \\
    \bottomrule
    \end{tabular}
    \caption{Language coverage of the datasets used for training.}
    \label{tab:training_dataset_languages}
\end{table}

\subsection{Datasets and Language Coverage}
We use MMARCO~\cite{mmarco}, RLHN~\cite{rlhn}, and MIRACL~\cite{miracl} to train LAMAR. MMARCO\footnote{\url{https://huggingface.co/datasets/unicamp-dl/mmarco}} is a multilingual version of the MS MARCO passage ranking dataset, and the processed dataset used in our experiments provides aligned query and positive document pairs across 14 languages. RLHN\footnote{\url{https://huggingface.co/datasets/rlhn/rlhn-250K}} is an English retrieval training collection obtained by using a cascading LLM procedure to identify and relabel false negatives in seven datasets from the BGE~\cite{bge} training collection. Each record associates a query with multiple positive documents, including relabeled hard negatives, as well as the remaining negative documents. MIRACL is a multilingual ad hoc retrieval dataset constructed from Wikipedia collections. For training, we use the MIRACL multilingual triplets dataset\footnote{\url{https://huggingface.co/datasets/nlpai-lab/miracl-multilingual-triplets}}, which is derived from 2,863 triplets in the English training split of MIRACL. The original English examples were translated into 50 additional languages with TranslationGemma-27B~\cite{translategemma}, producing 51 aligned language subsets in which examples sharing the same identifier correspond to the same source triplet. Each example contains a query, a positive document, and a negative document, allowing the semantic relation of a triplet to be retained across languages. Table~\ref{tab:training_dataset_languages} summarizes the language coverage of the three datasets.

\subsection{Data Construction for English-Anchored Relevance Distillation}
English-anchored relevance distillation pairs an English teacher input with a student input that represents the same underlying query--document relation. For a parallel dataset $\mathcal{X}$, $q_i^{\ell}$ denotes the realization of query identity $i$ in language $\ell$, and $d_j^{\ell}$ denotes the realization of document identity $j$ in language $\ell$. Accordingly, $q_i^{\mathrm{en}}$ and $q_i^{\ell_q}$ express the same query, while $d_j^{\mathrm{en}}$ and $d_j^{\ell_d}$ express the same selected document. The teacher and student inputs therefore preserve both identities $i$ and $j$, including when document $j$ is selected from an example other than query example $i$. Algorithm~\ref{alg:stage1_data} summarizes this construction.

\begin{algorithm}[H]
\caption{English-anchored distillation data synthesis}
\label{alg:stage1_data}
\small
\begin{algorithmic}[1]
\REQUIRE Parallel datasets $\mathcal{C}$ with aligned language realizations and ordered language lists $\{\mathcal{L}_{\mathcal{X}}\}$; English pairs $\mathcal{E}$
\ENSURE Distillation dataset $\mathcal{D}_1$
\STATE $\mathcal{D}_1 \gets [\,]$
\FOR{each parallel dataset $\mathcal{X} \in \mathcal{C}$}
    \STATE $c \gets 0$
    \FOR{each example $i \in \mathcal{X}$}
        \STATE $\mathcal{S}_{\mathcal{X}}(i) \gets \operatorname{SelectDocs}(\mathcal{X},i)$
        \FOR{each document identity $j \in \mathcal{S}_{\mathcal{X}}(i)$}
            \STATE $(\ell_q,\ell_d) \gets \operatorname{LangPair}(c,\mathcal{L}_{\mathcal{X}})$
            \STATE $x_T \gets (q_i^{\mathrm{en}},d_j^{\mathrm{en}})$
            \STATE $x_S \gets (q_i^{\ell_q},d_j^{\ell_d})$
            \STATE Append $(x_T,x_S)$ to $\mathcal{D}_1$
            \STATE $c \gets c+1$
        \ENDFOR
    \ENDFOR
\ENDFOR
\FOR{each $(q^{\mathrm{en}},d^{\mathrm{en}}) \in \mathcal{E}$}
    \STATE Append $\big((q^{\mathrm{en}},d^{\mathrm{en}}),(q^{\mathrm{en}},d^{\mathrm{en}})\big)$ to $\mathcal{D}_1$
\ENDFOR
\RETURN $\mathcal{D}_1$
\end{algorithmic}
\end{algorithm}

In Algorithm~\ref{alg:stage1_data}, the function $\operatorname{SelectDocs}(\mathcal{X},i)$ constructs the dataset-specific sequence $\mathcal{S}_{\mathcal{X}}(i)$ of document identities paired with query $i$. Documents drawn from other examples are randomly sampled without replacement within each sampling round, excluding the current example, using a fixed random seed. The function $\operatorname{LangPair}(c,\mathcal{L})$ assigns query and document languages by cycling through all pairs in the ordered Cartesian product $\mathcal{L}\times\mathcal{L}$, ensuring balanced coverage of every possible language combination within each dataset. For MMARCO, denoted by $\mathcal{M}$, $\mathcal{S}_{\mathcal{M}}(i)$ contains the document originally paired with query $i$ and three documents randomly sampled from distinct examples, yielding four query--document relations per query. The sampled documents are drawn from the positive fields of their source examples but are not treated as labeled positives for query $i$; their relevance to $q_i$ is instead represented by the teacher score. For MIRACL, we perform 20 sampling rounds for each triplet. In each round, the query is paired separately with its original positive and negative documents and with four documents randomly sampled from distinct examples, comprising two from positive fields and two from negative fields. This produces 120 distillation pairs per triplet. 

For RLHN-250K, we exclude the msmarco subset and pair each remaining query with every positive and negative document in its set. These query--document pairs constitute $\mathcal{E}$ in Algorithm~\ref{alg:stage1_data}. This procedure produces 6,665,428 pairs in total: 1,663,752 from MMARCO, 343,560 from MIRACL, and 4,658,116 from RLHN-250K.

\subsection{Data Construction for Preference Alignment for Language Coherence}
Algorithm~\ref{alg:stage2_data} presents the general listwise synthesis procedure used for preference alignment for language coherence.
\begin{algorithm}[H]
\caption{Language-coherence listwise data synthesis}
\label{alg:stage2_data}
\small
\begin{algorithmic}[1]
\REQUIRE Parallel triplet dataset $\mathcal{X}$ with ordered language list $\mathcal{L}_{\mathcal{X}}$; repetition factor $R$
\ENSURE Listwise dataset $\mathcal{D}_2$
\STATE $\mathcal{D}_2 \gets [\,]$; $c \gets 0$
\FOR{each example $i \in \mathcal{X}$}
    \FOR{$r=1$ to $R$}
        \STATE $(\mathrm{src},\mathrm{tgt}) \gets \operatorname{LangPair}(c,\mathcal{L}_{\mathcal{X}})$
        \STATE Append $(q_{i,\mathrm{src}},d_{i,\mathrm{src}}^{+},d_{i,\mathrm{tgt}}^{+},d_{i,\mathrm{src}}^{-},d_{i,\mathrm{tgt}}^{-})$ to $\mathcal{D}_2$
        \STATE $c \gets c+1$
    \ENDFOR
\ENDFOR
\RETURN $\mathcal{D}_2$
\end{algorithmic}
\end{algorithm}
For Stage 2, $\mathcal{X}$ is the MIRACL multilingual triplet dataset, and $R$ determines the number of language-pair assignments generated for each triplet. We set $R$ to 3 and cycle through all 2,601 source and target language pairs formed by the 51 languages. The assignments are balanced across these combinations, with each language pair appearing in three or four instances. For every assignment, the query and the source and target language versions of both the positive and negative documents share the same MIRACL row $i$; no documents from other rows are introduced. Because the Cartesian product includes pairs in the same language, $\mathrm{src}$ and $\mathrm{tgt}$ may denote the same or different languages. The resulting dataset contains 8,589 listwise instances.

\section{Full Results}
\subsection{Language-Coherence Evaluation}

The detailed results for BELEBELE are provided as follows:
\begin{itemize}
    \item \textbf{nDCG@1}: Table~\ref{tab:belebele_ndcg1}
    \item \textbf{nDCG@10}: Table~\ref{tab:belebele_ndcg10}
    \item \textbf{MRR@10}: Table~\ref{tab:belebele_mrr10}
\end{itemize}
\input{appendix/table/belebele_ndcg1}
\input{appendix/table/belebele_ndcg10}
\input{appendix/table/belebele_mrr}
The detailed results for XQuAD are provided as follows:
\begin{itemize}
    \item \textbf{nDCG@1}: Table~\ref{tab:xquad_ndcg1}
    \item \textbf{nDCG@10}: Table~\ref{tab:xquad_ndcg10}
    \item \textbf{MRR@10}: Table~\ref{tab:xquad_mrr10}
\end{itemize}
\input{appendix/table/xquad_ndcg1}
\input{appendix/table/xquad_ndcg10}

\input{appendix/table/xquad_mrr}
\subsection{Multilingual Reranking Evaluation}
The detailed results for the MTEB multilingual reranking benchmarks are provided as follows:
\begin{itemize}
    \item \textbf{XGLUE}: Table~\ref{tab:xglue_ndcg10}
    \item \textbf{HUME}: Table~\ref{tab:hume_ndcg10}
    \item \textbf{MLDR}: Table~\ref{tab:mldr_main_ndcg10}
    \item \textbf{WIKIPEDIA}: Table~\ref{tab:wiki_ndcg10}
\end{itemize}

\input{appendix/table/xglue}
\input{appendix/table/hume}

\input{appendix/table/mldr}
\input{appendix/table/wikipedia}

%% file: appendix/table/belebele_ndcg1.tex
\begin{table*}[t]
\centering
\small
\renewcommand{\arraystretch}{1.2}
\resizebox{\textwidth}{!}{
\begin{tabular}{lccccccccccccccc}
\toprule
\textbf{Model}
& \textbf{ar} & \textbf{de} & \textbf{en} & \textbf{es}
& \textbf{fr} & \textbf{hi} & \textbf{id} & \textbf{it}
& \textbf{ja} & \textbf{nl} & \textbf{pt} & \textbf{ru}
& \textbf{vi} & \textbf{zh} & \textbf{Avg.} \\
\midrule
gte-multilingual-reranker-base
& 79.44 & 80.89 & 85.56 & 65.89 & 71.33
& 76.00 & 82.89 & 76.56 & 79.22 & 59.00
& 56.00 & 77.44 & 93.89 & 94.56 & 77.05 \\

jina-reranker-v2-base-multilingual
& 53.89 & 82.33 & 69.89 & 57.89 & 54.22
& 44.11 & 51.33 & 50.44 & 59.11 & 62.44
& 48.22 & 55.89 & 84.56 & 72.22 & 60.47 \\

bge-reranker-v2-m3
& 85.78 & 89.67 & 65.33 & 83.11 & 91.89
& 78.67 & 88.78 & 83.00 & 82.78 & 83.00
& 83.22 & 90.67 & 98.22 & 83.44 & 84.83 \\

Qwen3-Reranker-0.6B
& 75.67 & 80.89 & 28.67 & 56.22 & 63.11
& 79.67 & 66.67 & 51.67 & 74.56 & 70.67
& 57.44 & 72.78 & 85.67 & 83.67 & 67.67 \\

jina-reranker-v3
& 88.22 & 63.11 & 29.67 & 55.00 & 75.56
& 93.78 & 92.22 & 70.11 & 93.11 & 90.44
& 65.89 & 94.67 & 96.11 & 85.44 & 78.10 \\

Prism-Qwen3.5-Reranker-0.8B
& 82.33 & 62.89 & 41.11 & 70.00 & 66.11
& 88.56 & 74.56 & 56.56 & 68.89 & 71.56
& 59.00 & 88.67 & 91.67 & 73.67 & 71.11 \\

llama-nemotron-rerank-1b-v2
& 98.00 & 97.00 & 87.89 & 93.44 & 96.11
& 92.22 & 98.44 & 90.78 & 93.78 & 88.56
& 89.67 & 96.89 & 99.22 & 98.44 & 94.32 \\

ctxl-rerank-v2-instruct-multilingual-1b
& 96.44 & 98.22 & 83.67 & 88.56 & 98.89
& 95.00 & 96.11 & 96.11 & 94.33 & 95.44
& 94.00 & 97.89 & 98.56 & 91.67 & 94.64 \\

Prism-Qwen3.5-Reranker-2B
& 64.00 & 57.33 & 38.78 & 53.33 & 57.89
& 84.89 & 57.33 & 57.67 & 59.33 & 57.56
& 44.56 & 68.89 & 75.67 & 71.44 & 60.62 \\

bge-reranker-v2-gemma
& 82.00 & 90.78 & 75.22 & 79.11 & 93.11
& 80.44 & 91.56 & 87.11 & 78.44 & 88.33
& 83.89 & 92.22 & 96.78 & 64.67 & 84.55 \\

Qwen3-Reranker-4B
& 41.67 & 50.44 & 37.44 & 58.00 & 38.44
& 42.89 & 38.00 & 28.33 & 37.44 & 40.44
& 35.33 & 46.00 & 59.56 & 52.33 & 43.31 \\

zerank-2-reranker
& 85.33 & 81.44 & 79.56 & 73.33 & 56.78
& 78.78 & 74.67 & 69.67 & 74.89 & 69.33
& 57.56 & 91.44 & 86.00 & 67.11 & 74.71 \\

Prism-Qwen3.5-Reranker-4B
& 74.89 & 64.78 & 53.44 & 69.11 & 63.44
& 85.78 & 51.00 & 63.00 & 58.78 & 67.00
& 51.56 & 85.44 & 77.22 & 77.00 & 67.32 \\

\midrule
\textbf{LAMAR (Ours)}
& 99.33 & 99.22 & 99.11 & 84.78 & 88.56
& 92.44 & 92.78 & 94.89 & 93.67 & 89.44
& 93.89 & 99.33 & 99.00 & 98.78 & 94.66 \\
\bottomrule
\end{tabular}
}
\caption{BELEBELE results measured by nDCG@1 across languages.}
\label{tab:belebele_ndcg1}
\end{table*}

%% file: appendix/table/belebele_ndcg10.tex
\begin{table*}[t]
\centering
\small
\renewcommand{\arraystretch}{1.2}
\resizebox{\textwidth}{!}{
\begin{tabular}{lccccccccccccccc}
\toprule
\textbf{Model}
& \textbf{ar} & \textbf{de} & \textbf{en} & \textbf{es}
& \textbf{fr} & \textbf{hi} & \textbf{id} & \textbf{it}
& \textbf{ja} & \textbf{nl} & \textbf{pt} & \textbf{ru}
& \textbf{vi} & \textbf{zh} & \textbf{Avg.} \\
\midrule
gte-multilingual-reranker-base
& 89.13 & 90.59 & 93.36 & 83.18 & 86.84
& 87.01 & 91.80 & 88.86 & 89.24 & 77.95
& 78.97 & 88.36 & 97.27 & 97.50 & 88.58 \\

jina-reranker-v2-base-multilingual
& 75.97 & 92.01 & 84.24 & 79.11 & 76.79
& 70.90 & 74.66 & 73.74 & 78.74 & 82.18
& 74.92 & 75.64 & 92.83 & 86.57 & 79.88 \\

bge-reranker-v2-m3
& 92.64 & 95.15 & 81.63 & 92.55 & 96.24
& 89.30 & 94.52 & 92.07 & 90.88 & 92.03
& 93.23 & 95.34 & 99.19 & 92.27 & 92.65 \\

Qwen3-Reranker-0.6B
& 87.35 & 90.64 & 56.42 & 78.00 & 81.97
& 87.11 & 81.99 & 74.92 & 86.70 & 84.44
& 79.23 & 85.65 & 93.19 & 92.27 & 82.85 \\

jina-reranker-v3
& 93.59 & 82.93 & 58.89 & 76.03 & 87.48
& 96.55 & 96.13 & 83.48 & 96.74 & 95.42
& 82.07 & 97.45 & 98.10 & 93.19 & 88.43 \\

Prism-Qwen3.5-Reranker-0.8B
& 91.72 & 81.76 & 68.88 & 85.57 & 83.74
& 94.01 & 86.88 & 78.70 & 84.55 & 85.83
& 81.32 & 94.75 & 96.34 & 87.29 & 85.81 \\

llama-nemotron-rerank-1b-v2
& 99.02 & 98.67 & 94.02 & 96.85 & 98.23
& 95.50 & 99.18 & 95.65 & 96.90 & 94.58
& 95.16 & 98.45 & 99.64 & 99.14 & 97.21 \\

ctxl-rerank-v2-instruct-multilingual-1b
& 98.24 & 99.25 & 92.00 & 94.59 & 99.53
& 97.82 & 98.25 & 98.30 & 97.62 & 98.07
& 97.34 & 99.08 & 99.36 & 96.04 & 97.54 \\

Prism-Qwen3.5-Reranker-2B
& 81.18 & 78.27 & 66.34 & 75.71 & 78.30
& 91.73 & 77.89 & 78.11 & 78.44 & 77.52
& 72.19 & 84.55 & 88.12 & 86.27 & 79.62 \\

bge-reranker-v2-gemma
& 90.31 & 95.84 & 88.21 & 89.87 & 97.04
& 89.44 & 95.97 & 94.10 & 89.51 & 94.43
& 92.87 & 96.18 & 98.50 & 82.62 & 92.49 \\

Qwen3-Reranker-4B
& 64.14 & 73.10 & 62.26 & 76.42 & 65.87
& 62.75 & 63.55 & 57.96 & 62.15 & 66.28
& 64.23 & 69.21 & 78.61 & 71.71 & 67.02 \\

zerank-2-reranker
& 92.66 & 90.87 & 90.16 & 86.77 & 77.75
& 88.56 & 86.83 & 85.25 & 86.58 & 84.19
& 79.70 & 95.74 & 92.60 & 82.64 & 87.16 \\

Prism-Qwen3.5-Reranker-4B
& 86.85 & 81.91 & 75.92 & 84.62 & 81.84
& 92.37 & 73.16 & 81.00 & 77.74 & 83.28
& 76.34 & 93.16 & 88.54 & 88.47 & 83.23 \\

\midrule
\textbf{LAMAR (Ours)}
& 99.75 & 99.70 & 99.65 & 93.24 & 94.64
& 96.17 & 96.76 & 97.85 & 96.87 & 95.40
& 97.51 & 99.72 & 99.62 & 99.48 & 97.60 \\
\bottomrule
\end{tabular}
}
\caption{BELEBELE results measured by nDCG@10 across languages.}
\label{tab:belebele_ndcg10}
\end{table*}

%% file: appendix/table/belebele_mrr.tex
\begin{table*}[t]
\centering
\small
\renewcommand{\arraystretch}{1.2}
\resizebox{\textwidth}{!}{
\begin{tabular}{lccccccccccccccc}
\toprule
\textbf{Model}
& \textbf{ar} & \textbf{de} & \textbf{en} & \textbf{es}
& \textbf{fr} & \textbf{hi} & \textbf{id} & \textbf{it}
& \textbf{ja} & \textbf{nl} & \textbf{pt} & \textbf{ru}
& \textbf{vi} & \textbf{zh} & \textbf{Avg.} \\
\midrule
gte-multilingual-reranker-base
& 87.50 & 88.97 & 92.31 & 77.52 & 83.33
& 83.37 & 89.53 & 85.04 & 85.78 & 70.79
& 70.86 & 83.89 & 95.84 & 96.40 & 85.08 \\

jina-reranker-v2-base-multilingual
& 70.09 & 89.95 & 80.23 & 72.13 & 69.27
& 62.14 & 66.56 & 65.49 & 71.83 & 75.60
& 65.95 & 67.70 & 89.93 & 81.97 & 73.49 \\

bge-reranker-v2-m3
& 91.41 & 95.53 & 78.91 & 89.67 & 95.48
& 85.85 & 92.14 & 89.45 & 88.27 & 88.37
& 89.83 & 92.69 & 98.67 & 88.17 & 90.32 \\

Qwen3-Reranker-0.6B
& 85.02 & 88.83 & 47.15 & 70.69 & 76.96
& 84.70 & 77.01 & 67.14 & 82.52 & 79.19
& 71.35 & 80.38 & 89.58 & 88.42 & 77.78 \\

jina-reranker-v3
& 91.92 & 77.44 & 47.91 & 68.97 & 83.71
& 95.75 & 94.98 & 79.06 & 95.71 & 93.95
& 76.75 & 96.62 & 97.54 & 91.04 & 85.10 \\

Prism-Qwen3.5-Reranker-0.8B
& 94.78 & 88.50 & 77.45 & 75.36 & 80.12
& 91.02 & 80.24 & 67.16 & 75.28 & 75.24
& 64.07 & 87.06 & 88.65 & 70.82 & 79.70 \\

llama-nemotron-rerank-1b-v2
& 98.80 & 98.27 & 92.38 & 95.88 & 97.65
& 94.45 & 98.85 & 94.17 & 96.00 & 92.79
& 93.58 & 97.97 & 99.52 & 98.90 & 96.37 \\

ctxl-rerank-v2-instruct-multilingual-1b
& 97.73 & 98.99 & 89.51 & 92.89 & 99.37
& 97.08 & 97.73 & 97.75 & 96.80 & 97.44
& 96.45 & 98.76 & 99.14 & 94.46 & 96.72 \\

Prism-Qwen3.5-Reranker-2B
& 87.13 & 87.05 & 72.68 & 60.71 & 72.80
& 88.17 & 69.22 & 65.44 & 68.29 & 59.47
& 48.90 & 65.12 & 74.59 & 64.24 & 70.27 \\

bge-reranker-v2-gemma
& 87.99 & 94.66 & 85.40 & 86.61 & 96.15
& 86.35 & 94.53 & 91.97 & 85.72 & 92.34
& 89.89 & 94.61 & 97.96 & 76.31 & 90.04 \\

Qwen3-Reranker-4B
& 58.72 & 67.83 & 54.29 & 69.99 & 57.52
& 56.51 & 54.46 & 47.16 & 53.54 & 56.57
& 54.01 & 59.55 & 70.52 & 62.43 & 58.79 \\

zerank-2-reranker
& 90.93 & 88.36 & 87.40 & 82.46 & 71.60
& 85.67 & 82.91 & 80.72 & 82.43 & 79.32
& 72.77 & 94.38 & 89.76 & 76.42 & 83.22 \\

Prism-Qwen3.5-Reranker-4B
& 93.68 & 89.96 & 82.18 & 74.99 & 79.76
& 89.84 & 66.41 & 72.43 & 69.35 & 68.80
& 60.21 & 86.25 & 75.78 & 67.70 & 76.95 \\

\midrule
\textbf{LAMAR (Ours)}
& 99.67 & 99.65 & 99.53 & 90.85 & 93.09
& 94.90 & 95.84 & 97.06 & 95.88 & 93.69
& 96.49 & 99.56 & 99.48 & 99.31 & 96.79 \\
\bottomrule
\end{tabular}
}
\caption{BELEBELE results measured by MRR@10 across languages.}
\label{tab:belebele_mrr10}
\end{table*}

%% file: appendix/table/xquad_ndcg1.tex
\begin{table*}[t]
\centering
\small
\renewcommand{\arraystretch}{1.2}
\resizebox{0.9\textwidth}{!}{
\begin{tabular}{lccccccccccccc}
\toprule
\textbf{Model}
& \textbf{ar} & \textbf{de} & \textbf{el} & \textbf{en}
& \textbf{es} & \textbf{hi} & \textbf{ro} & \textbf{ru}
& \textbf{th} & \textbf{tr} & \textbf{vi} & \textbf{zh}
& \textbf{Avg.} \\
\midrule
gte-multilingual-reranker-base
& 83.36 & 74.54 & 84.79 & 66.22 & 80.42 & 85.80
& 84.62 & 81.76 & 82.18 & 85.29 & 90.34 & 91.76 & 82.59 \\

jina-reranker-v2-base-multilingual
& 39.58 & 63.11 & 63.36 & 50.08 & 49.41 & 36.13
& 73.95 & 43.11 & 62.27 & 82.86 & 63.19 & 60.59 & 57.30 \\

bge-reranker-v2-m3
& 88.23 & 85.04 & 88.82 & 32.94 & 86.81 & 87.48
& 93.45 & 86.56 & 87.81 & 96.56 & 94.79 & 87.56 & 84.67 \\

Qwen3-Reranker-0.6B
& 67.73 & 68.49 & 59.41 & 26.39 & 60.17 & 59.24
& 57.82 & 61.51 & 60.34 & 76.98 & 69.83 & 72.77 & 61.72 \\

jina-reranker-v3
& 80.42 & 60.50 & 93.45 & 27.23 & 56.05 & 94.03
& 81.43 & 78.82 & 97.73 & 90.17 & 85.21 & 87.90 & 77.75 \\

Prism-Qwen3.5-Reranker-0.8B
& 75.97 & 56.30 & 73.70 & 44.96 & 62.02 & 83.19
& 79.66 & 82.18 & 86.30 & 73.19 & 79.41 & 70.50 & 72.28 \\

llama-nemotron-rerank-1b-v2
& 95.04 & 96.47 & 94.79 & 84.96 & 97.81 & 96.13
& 97.23 & 96.30 & 97.90 & 97.90 & 97.06 & 98.40 & 95.83 \\

ctxl-rerank-v2-instruct-multilingual-1b
& 85.13 & 92.27 & 83.95 & 87.65 & 88.74 & 86.98
& 91.85 & 88.15 & 88.91 & 86.47 & 93.03 & 92.35 & 88.79 \\

Prism-Qwen3.5-Reranker-2B
& 70.50 & 61.09 & 72.10 & 54.12 & 55.88 & 83.28
& 75.38 & 69.16 & 85.97 & 70.42 & 75.63 & 71.43 & 70.41 \\

bge-reranker-v2-gemma
& 84.87 & 89.33 & 88.91 & 68.07 & 91.26 & 89.92
& 88.82 & 88.57 & 90.42 & 93.36 & 94.71 & 71.26 & 86.63 \\

Qwen3-Reranker-4B
& 26.47 & 39.66 & 30.92 & 20.17 & 35.21 & 35.38
& 28.40 & 25.80 & 33.95 & 43.03 & 42.60 & 34.79 & 33.03 \\

zerank-2-reranker
& 66.89 & 60.67 & 62.86 & 64.12 & 60.34 & 57.39
& 61.85 & 70.50 & 71.68 & 60.84 & 68.24 & 52.19 & 63.13 \\

Prism-Qwen3.5-Reranker-4B
& 62.94 & 53.53 & 74.29 & 63.53 & 62.35 & 83.36
& 76.98 & 74.96 & 84.20 & 64.71 & 72.60 & 72.86 & 70.53 \\

\midrule
\textbf{LAMAR (Ours)}
& 98.40 & 98.82 & 96.98 & 98.49 & 94.96 & 95.55
& 94.45 & 98.74 & 93.03 & 98.40 & 97.48 & 97.39 & 96.89 \\
\bottomrule
\end{tabular}
}
\caption{XQuAD retrieval results measured by nDCG@1 across languages.}
\label{tab:xquad_ndcg1}
\end{table*}

%% file: appendix/table/xquad_ndcg10.tex
\begin{table*}[t]
\centering
\small
\renewcommand{\arraystretch}{1.2}
\resizebox{0.9\textwidth}{!}{
\begin{tabular}{lccccccccccccc}
\toprule
\textbf{Model}
& \textbf{ar} & \textbf{de} & \textbf{el} & \textbf{en}
& \textbf{es} & \textbf{hi} & \textbf{ro} & \textbf{ru}
& \textbf{th} & \textbf{tr} & \textbf{vi} & \textbf{zh}
& \textbf{Avg.} \\
\midrule
gte-multilingual-reranker-base
& 91.20 & 87.37 & 92.49 & 84.06 & 91.33 & 92.76
& 92.89 & 91.21 & 90.84 & 92.64 & 95.46 & 95.91 & 91.51 \\

jina-reranker-v2-base-multilingual
& 68.27 & 82.82 & 82.86 & 74.68 & 77.08 & 68.04
& 88.71 & 71.36 & 81.70 & 92.41 & 82.58 & 81.77 & 79.36 \\

bge-reranker-v2-m3
& 94.15 & 92.51 & 94.69 & 61.66 & 94.08 & 94.05
& 97.12 & 93.61 & 94.26 & 98.56 & 97.69 & 94.06 & 92.20 \\

Qwen3-Reranker-0.6B
& 82.95 & 83.96 & 73.81 & 59.51 & 80.82 & 74.36
& 78.78 & 80.78 & 79.13 & 88.59 & 85.21 & 87.01 & 79.58 \\

jina-reranker-v3
& 89.22 & 80.37 & 96.50 & 59.37 & 77.46 & 96.97
& 90.11 & 89.13 & 98.90 & 95.11 & 92.47 & 94.10 & 88.31 \\

Prism-Qwen3.5-Reranker-0.8B
& 88.78 & 78.73 & 87.42 & 73.38 & 82.84 & 91.08
& 90.67 & 91.61 & 92.75 & 86.69 & 90.59 & 86.30 & 86.74 \\

llama-nemotron-rerank-1b-v2
& 97.51 & 98.51 & 97.56 & 92.71 & 99.08 & 97.88
& 98.84 & 98.39 & 98.96 & 98.96 & 98.64 & 99.23 & 98.02 \\

ctxl-rerank-v2-instruct-multilingual-1b
& 93.36 & 96.68 & 92.08 & 94.34 & 94.82 & 94.53
& 96.24 & 94.74 & 95.08 & 93.82 & 96.87 & 96.47 & 94.92 \\

Prism-Qwen3.5-Reranker-2B
& 85.23 & 81.18 & 86.37 & 77.86 & 78.75 & 89.78
& 87.98 & 84.94 & 92.62 & 84.78 & 88.24 & 86.30 & 85.34 \\

bge-reranker-v2-gemma
& 91.89 & 94.73 & 94.60 & 84.36 & 96.13 & 94.79
& 94.95 & 94.64 & 95.08 & 96.63 & 97.39 & 85.79 & 93.42 \\

Qwen3-Reranker-4B
& 53.10 & 66.54 & 48.77 & 51.81 & 63.30 & 53.70
& 57.85 & 55.40 & 58.31 & 67.57 & 68.44 & 60.93 & 58.81 \\

zerank-2-reranker
& 82.46 & 80.08 & 73.97 & 82.73 & 80.59 & 73.51
& 80.80 & 85.04 & 84.89 & 78.43 & 83.72 & 73.48 & 79.98 \\

Prism-Qwen3.5-Reranker-4B
& 80.63 & 76.13 & 87.61 & 82.26 & 82.06 & 90.66
& 89.13 & 87.78 & 92.17 & 80.93 & 86.81 & 86.08 & 85.19 \\

\midrule
\textbf{LAMAR (Ours)}
& 99.20 & 99.52 & 98.52 & 99.38 & 97.98 & 97.79
& 97.60 & 99.42 & 96.70 & 99.20 & 98.99 & 98.72 & 98.59 \\
\bottomrule
\end{tabular}
}
\caption{XQuAD retrieval results measured by nDCG@10 across languages.}
\label{tab:xquad_ndcg10}
\end{table*}

%% file: appendix/table/xquad_mrr.tex
\begin{table*}[t]
\centering
\small
\renewcommand{\arraystretch}{1.2}
\resizebox{0.9\textwidth}{!}{
\begin{tabular}{lccccccccccccc}
\toprule
\textbf{Model}
& \textbf{ar} & \textbf{de} & \textbf{el} & \textbf{en}
& \textbf{es} & \textbf{hi} & \textbf{ro} & \textbf{ru}
& \textbf{th} & \textbf{tr} & \textbf{vi} & \textbf{zh}
& \textbf{Avg.} \\
\midrule
gte-multilingual-reranker-base
& 89.30 & 85.52 & 90.90 & 79.95 & 89.05 & 90.53
& 89.02 & 87.97 & 87.60 & 89.72 & 92.94 & 94.22 & 88.89 \\

jina-reranker-v2-base-multilingual
& 59.59 & 78.02 & 77.72 & 67.57 & 68.68 & 57.39
& 84.46 & 61.29 & 75.29 & 89.44 & 76.04 & 75.11 & 72.55 \\

bge-reranker-v2-m3
& 93.31 & 93.04 & 94.44 & 53.60 & 94.71 & 91.27
& 95.52 & 89.51 & 90.77 & 97.61 & 96.32 & 88.57 & 89.89 \\

Qwen3-Reranker-0.6B
& 78.81 & 80.13 & 68.73 & 47.96 & 74.47 & 68.95
& 71.43 & 74.23 & 72.34 & 84.58 & 79.56 & 81.56 & 73.56 \\

jina-reranker-v3
& 86.41 & 74.18 & 95.53 & 47.39 & 70.55 & 96.09
& 87.11 & 85.73 & 98.54 & 93.57 & 90.11 & 92.24 & 84.79 \\

Prism-Qwen3.5-Reranker-0.8B
& 89.28 & 81.22 & 87.62 & 73.45 & 75.76 & 87.54
& 84.57 & 84.43 & 89.32 & 78.41 & 82.55 & 72.42 & 82.21 \\

llama-nemotron-rerank-1b-v2
& 96.80 & 98.00 & 96.80 & 90.23 & 98.77 & 97.38
& 98.53 & 97.85 & 98.64 & 98.63 & 98.15 & 98.93 & 97.39 \\

ctxl-rerank-v2-instruct-multilingual-1b
& 91.31 & 95.59 & 89.80 & 92.56 & 93.12 & 92.67
& 95.03 & 92.92 & 93.44 & 91.66 & 95.86 & 95.25 & 93.27 \\

Prism-Qwen3.5-Reranker-2B
& 87.42 & 85.14 & 86.93 & 74.97 & 68.01 & 86.79
& 78.09 & 73.46 & 89.27 & 74.93 & 78.44 & 71.45 & 79.58 \\

bge-reranker-v2-gemma
& 89.82 & 93.39 & 93.16 & 79.89 & 94.62 & 92.98
& 92.97 & 92.55 & 93.47 & 95.42 & 96.32 & 80.24 & 91.24 \\

Qwen3-Reranker-4B
& 43.78 & 58.40 & 41.84 & 40.08 & 53.37 & 46.31
& 45.49 & 42.70 & 47.93 & 57.42 & 57.19 & 48.86 & 48.61 \\

zerank-2-reranker
& 77.57 & 74.79 & 69.85 & 77.60 & 74.34 & 67.92
& 74.56 & 80.07 & 80.21 & 71.57 & 77.99 & 65.21 & 74.31 \\

Prism-Qwen3.5-Reranker-4B
& 84.18 & 80.90 & 89.29 & 83.16 & 76.55 & 88.12
& 83.65 & 79.22 & 87.98 & 67.58 & 74.98 & 72.05 & 80.64 \\

\midrule
\textbf{LAMAR (Ours)}
& 98.99 & 99.37 & 98.06 & 99.16 & 97.17 & 97.16
& 96.71 & 99.22 & 95.62 & 98.93 & 98.60 & 98.21 & 98.10 \\
\bottomrule
\end{tabular}
}
\caption{XQuAD retrieval results measured by MRR@10 across languages.}
\label{tab:xquad_mrr10}
\end{table*}

%% file: appendix/table/xglue.tex
\begin{table*}[t]
\centering
\small
\renewcommand{\arraystretch}{1.2}
\resizebox{0.65\textwidth}{!}{
\begin{tabular}{lcccccccc}
\toprule
\textbf{Model}
& \textbf{de} & \textbf{en} & \textbf{es} & \textbf{fr}
& \textbf{it} & \textbf{pt} & \textbf{zh} & \textbf{Avg.} \\
\midrule
gte-multilingual-reranker-base
& 78.82 & 78.12 & 77.07 & 76.78
& 74.22 & 77.27 & 71.59 & 76.27 \\

jina-reranker-v2-base-multilingual
& 79.48 & 78.66 & 77.47 & 77.08
& 75.50 & 76.83 & 72.38 & 76.77 \\

bge-reranker-v2-m3
& 79.45 & 77.32 & 77.65 & 76.42
& 74.98 & 77.34 & 72.29 & 76.49 \\

Qwen3-Reranker-0.6B
& 79.23 & 78.49 & 77.98 & 77.03
& 74.75 & 76.39 & 71.77 & 76.52 \\

jina-reranker-v3
& 82.67 & 81.79 & 81.49 & 80.59
& 79.43 & 81.43 & 77.46 & 80.69 \\

Prism-Qwen3.5-Reranker-0.8B
& 77.18 & 76.08 & 75.72 & 74.39
& 71.53 & 74.94 & 68.95 & 74.11 \\

llama-nemotron-rerank-1b-v2
& 80.11 & 79.37 & 78.63 & 78.28
& 77.11 & 78.86 & 73.27 & 77.95 \\

ctxl-rerank-v2-instruct-multilingual-1b
& 80.02 & 78.96 & 79.16 & 78.31
& 76.67 & 78.54 & 73.00 & 77.81 \\

Prism-Qwen3.5-Reranker-2B
& 77.52 & 76.42 & 76.17 & 74.99
& 72.89 & 76.48 & 69.15 & 74.80 \\

bge-reranker-v2-gemma
& 79.38 & 78.28 & 77.45 & 76.75
& 75.77 & 77.35 & 72.29 & 76.75 \\

Qwen3-Reranker-4B
& 79.54 & 78.28 & 77.49 & 76.72
& 75.30 & 76.98 & 70.50 & 76.40 \\

zerank-2-reranker
& 80.31 & 79.09 & 79.83 & 78.26
& 76.91 & 79.06 & 72.59 & 78.01 \\

Prism-Qwen3.5-Reranker-4B
& 78.30 & 76.58 & 76.65 & 75.33
& 72.87 & 76.09 & 68.87 & 74.96 \\

\midrule
\textbf{LAMAR (Ours)}
& 79.69 & 78.45 & 78.17 & 77.04
& 75.41 & 77.50 & 72.93 & 77.03 \\
\bottomrule
\end{tabular}
}
\caption{XGLUE retrieval results measured by nDCG@10 across languages.}
\label{tab:xglue_ndcg10}
\end{table*}

%% file: appendix/table/hume.tex
\begin{table*}[t]
\centering
\small
\renewcommand{\arraystretch}{1.2}
\resizebox{0.45\textwidth}{!}{
\begin{tabular}{lcccc}
\toprule
\textbf{Model}
& \textbf{en} & \textbf{da} & \textbf{no} & \textbf{Avg.} \\
\midrule
gte-multilingual-reranker-base
& 92.98 & 95.50 & 94.21 & 94.23 \\

jina-reranker-v2-base-multilingual
& 90.70 & 94.64 & 94.88 & 93.41 \\

bge-reranker-v2-m3
& 90.72 & 96.31 & 96.62 & 94.55 \\

Qwen3-Reranker-0.6B
& 93.41 & 96.31 & 94.07 & 94.60 \\

jina-reranker-v3
& 92.18 & 97.54 & 94.65 & 94.79 \\

Prism-Qwen3.5-Reranker-0.8B
& 95.08 & 95.21 & 92.06 & 94.11 \\

llama-nemotron-rerank-1b-v2
& 94.41 & 98.77 & 94.88 & 96.02 \\

ctxl-rerank-v2-instruct-multilingual-1b
& 90.52 & 96.31 & 93.97 & 93.60 \\

Prism-Qwen3.5-Reranker-2B
& 95.87 & 95.06 & 94.52 & 95.15 \\

bge-reranker-v2-gemma
& 91.08 & 97.54 & 94.02 & 94.21 \\

Qwen3-Reranker-4B
& 93.04 & 98.77 & 95.39 & 95.73 \\

zerank-2-reranker
& 95.87 & 95.64 & 96.87 & 96.13 \\

Prism-Qwen3.5-Reranker-4B
& 94.64 & 94.21 & 95.50 & 94.78 \\

\midrule
\textbf{LAMAR (Ours)}
& 92.62 & 97.54 & 95.73 & 95.30 \\
\bottomrule
\end{tabular}
}
\caption{HUME retrieval results measured by nDCG@10 across languages.}
\label{tab:hume_ndcg10}
\end{table*}

%% file: appendix/table/mldr.tex
\begin{table*}[t]
\centering
\small
\renewcommand{\arraystretch}{1.2}
\resizebox{\textwidth}{!}{
\begin{tabular}{lcccccccccccccc}
\toprule
\textbf{Model}
& \textbf{ar} & \textbf{de} & \textbf{en} & \textbf{es}
& \textbf{fr} & \textbf{hi} & \textbf{it} & \textbf{ja}
& \textbf{ko} & \textbf{pt} & \textbf{ru} & \textbf{th}
& \textbf{zh} & \textbf{Avg.} \\
\midrule
gte-multilingual-reranker-base
& 99.45 & 99.38 & 98.45 & 100.00 & 99.75 & 97.27
& 99.82 & 99.20 & 98.45 & 99.10 & 99.57 & 99.20
& 94.87 & 98.81 \\

jina-reranker-v2-base-multilingual
& 86.39 & 89.77 & 95.62 & 94.87 & 94.32 & 77.54
& 88.18 & 84.59 & 82.45 & 91.96 & 92.46 & 85.00
& 57.08 & 86.17 \\

bge-reranker-v2-m3
& 97.54 & 98.48 & 97.59 & 99.75 & 98.98 & 93.70
& 98.76 & 97.56 & 96.14 & 98.85 & 99.16 & 99.45
& 90.19 & 97.40 \\

Qwen3-Reranker-0.6B
& 99.26 & 99.13 & 99.63 & 100.00 & 100.00 & 97.30
& 99.57 & 98.26 & 96.12 & 98.61 & 100.00 & 99.26
& 97.92 & 98.85 \\

jina-reranker-v3
& 91.09 & 96.10 & 95.53 & 97.47 & 96.63 & 84.56
& 93.89 & 91.43 & 90.19 & 96.07 & 93.98 & 94.73
& 92.63 & 93.41 \\

Prism-Qwen3.5-Reranker-0.8B
& 97.33 & 96.00 & 99.22 & 99.35 & 99.05 & 89.53
& 98.31 & 94.51 & 92.69 & 98.73 & 99.38 & 92.20
& 96.23 & 96.35 \\

llama-nemotron-rerank-1b-v2
& 94.95 & 97.37 & 98.45 & 99.49 & 98.91 & 90.81
& 97.92 & 94.21 & 93.17 & 98.43 & 97.23 & 98.64
& 93.86 & 96.42 \\

ctxl-rerank-v2-instruct-multilingual-1b
& 97.48 & 96.72 & 99.50 & 99.37 & 99.02 & 91.86
& 98.72 & 95.98 & 96.59 & 98.16 & 99.72 & 95.61
& 97.09 & 97.37 \\

Prism-Qwen3.5-Reranker-2B
& 96.61 & 95.71 & 99.09 & 99.66 & 99.06 & 90.33
& 98.54 & 94.83 & 92.80 & 97.93 & 99.63 & 93.86
& 95.96 & 96.46 \\

bge-reranker-v2-gemma
& 85.51 & 95.69 & 85.81 & 91.36 & 96.23 & 82.96
& 93.71 & 87.78 & 88.08 & 96.36 & 89.87 & 97.62
& 66.90 & 89.07 \\

Qwen3-Reranker-4B
& 99.57 & 100.00 & 99.85 & 100.00 & 99.82 & 97.56
& 99.75 & 99.82 & 98.39 & 99.28 & 100.00 & 99.20
& 98.84 & 99.39 \\

zerank-2-reranker
& 99.07 & 97.88 & 99.42 & 99.28 & 99.63 & 93.31
& 99.32 & 97.10 & 96.43 & 98.97 & 99.82 & 96.84
& 97.39 & 98.03 \\

Prism-Qwen3.5-Reranker-4B
& 98.15 & 96.64 & 99.37 & 99.75 & 99.49 & 93.66
& 99.20 & 96.08 & 94.70 & 98.76 & 99.82 & 93.25
& 96.29 & 97.32 \\

\midrule
\textbf{LAMAR (Ours)}
& 97.80 & 97.70 & 98.84 & 99.82 & 99.47 & 94.66
& 99.04 & 97.06 & 94.46 & 98.64 & 98.17 & 97.92
& 95.25 & 97.60 \\
\bottomrule
\end{tabular}
}
\caption{MLDR retrieval results measured by Main-nDCG@10 across languages.}
\label{tab:mldr_main_ndcg10}
\end{table*}

%% file: appendix/table/wikipedia.tex
\begin{table*}[t]
\centering
\small
\renewcommand{\arraystretch}{1.2}
\resizebox{\textwidth}{!}{
\begin{tabular}{lccccccccccccccccc}
\toprule
\textbf{Model}
& \textbf{bg} & \textbf{bn} & \textbf{cs} & \textbf{da}
& \textbf{de} & \textbf{en} & \textbf{fa} & \textbf{fi}
& \textbf{hi} & \textbf{it} & \textbf{nl} & \textbf{pt}
& \textbf{ro} & \textbf{sr} & \textbf{no} & \textbf{sv}
& \textbf{Avg.} \\
\midrule
gte-multilingual-reranker-base
& 93.29 & 88.12 & 92.81 & 91.88 & 92.39 & 94.49
& 92.33 & 93.65 & 91.19 & 92.84 & 92.23 & 92.44
& 92.43 & 90.54 & 90.01 & 93.20 & 92.11 \\

jina-reranker-v2-base-multilingual
& 94.01 & 92.45 & 94.57 & 94.38 & 94.17 & 94.09
& 93.70 & 95.39 & 92.56 & 93.57 & 94.08 & 93.79
& 94.03 & 94.46 & 92.89 & 95.25 & 93.96 \\

bge-reranker-v2-m3
& 94.33 & 91.63 & 95.09 & 94.60 & 93.94 & 95.24
& 93.77 & 95.08 & 92.36 & 94.28 & 93.85 & 93.28
& 93.62 & 94.26 & 92.91 & 95.12 & 93.96 \\

Qwen3-Reranker-0.6B
& 95.71 & 90.17 & 96.22 & 95.43 & 94.88 & 95.84
& 94.25 & 94.87 & 91.05 & 96.15 & 95.37 & 95.05
& 95.18 & 95.53 & 93.67 & 95.71 & 94.69 \\

jina-reranker-v3
& 94.92 & 92.16 & 96.09 & 95.88 & 95.34 & 96.69
& 93.56 & 95.29 & 92.60 & 95.64 & 95.24 & 94.93
& 94.99 & 94.75 & 93.84 & 95.34 & 94.83 \\

Prism-Qwen3.5-Reranker-0.8B
& 93.88 & 85.31 & 95.13 & 94.80 & 94.29 & 95.72
& 92.36 & 94.24 & 82.66 & 93.91 & 94.27 & 94.20
& 93.34 & 93.54 & 92.69 & 94.40 & 92.80 \\

llama-nemotron-rerank-1b-v2
& 94.69 & 89.94 & 95.42 & 95.57 & 95.09 & 96.69
& 94.02 & 95.19 & 92.28 & 95.35 & 94.68 & 94.16
& 94.30 & 95.19 & 93.66 & 95.82 & 94.50 \\

ctxl-rerank-v2-instruct-multilingual-1b
& 94.31 & 88.73 & 94.87 & 94.79 & 94.69 & 96.18
& 92.82 & 94.37 & 89.94 & 94.20 & 94.11 & 93.98
& 94.05 & 94.00 & 92.30 & 95.10 & 93.65 \\

Prism-Qwen3.5-Reranker-2B
& 94.92 & 87.96 & 95.99 & 95.22 & 94.71 & 96.23
& 92.55 & 94.75 & 87.61 & 94.89 & 94.55 & 94.71
& 93.80 & 94.15 & 93.39 & 95.18 & 93.79 \\

bge-reranker-v2-gemma
& 94.99 & 92.72 & 95.54 & 95.50 & 94.80 & 96.14
& 94.21 & 95.42 & 93.31 & 94.87 & 94.47 & 93.96
& 93.91 & 95.25 & 93.58 & 95.39 & 94.63 \\

Qwen3-Reranker-4B
& 97.03 & 92.25 & 97.78 & 97.40 & 96.75 & 97.10
& 96.13 & 97.09 & 92.40 & 97.61 & 96.39 & 96.51
& 96.75 & 96.75 & 95.83 & 97.20 & 96.31 \\

zerank-2-reranker
& 95.54 & 91.01 & 95.73 & 95.68 & 94.92 & 96.42
& 93.43 & 95.81 & 91.30 & 94.80 & 94.61 & 94.60
& 94.03 & 94.65 & 94.25 & 95.42 & 94.51 \\

Prism-Qwen3.5-Reranker-4B
& 95.60 & 88.92 & 95.95 & 95.50 & 95.18 & 96.60
& 92.76 & 95.81 & 88.81 & 95.09 & 94.62 & 94.62
& 94.39 & 94.84 & 94.17 & 95.63 & 94.28 \\

\midrule
\textbf{LAMAR (Ours)}
& 94.85 & 92.83 & 95.47 & 95.54 & 94.75 & 95.69
& 94.26 & 96.05 & 93.32 & 95.20 & 94.67 & 94.47
& 94.46 & 95.18 & 94.19 & 95.72 & 94.79 \\
\bottomrule
\end{tabular}
}
\caption{Wiki retrieval results measured by nDCG@10 across languages.}
\label{tab:wiki_ndcg10}
\end{table*}